\documentclass[sn-nature,Numbered,iicol,12pt]{sn-jnl}

\usepackage{tgtermes} 
\usepackage[superscript]{cite}

\usepackage{graphicx}%
\usepackage{multirow}%
\usepackage{amsmath,amssymb,amsfonts}%
\usepackage{amsthm}%
\usepackage{mathrsfs}%
\usepackage[title]{appendix}%
\usepackage{xcolor}%
\usepackage{textcomp}%
\usepackage{manyfoot}%
\usepackage{booktabs}%
\usepackage{listings}%
\usepackage{placeins}
\usepackage{float}
\usepackage{comment}

\theoremstyle{thmstyleone}%
\theoremstyle{thmstyletwo}%

\theoremstyle{thmstylethree}%

\begin{document}
	
\title[First demonstration of Super-X divertor exhaust control for transient heat load management in compact fusion reactors]{First demonstration of Super-X divertor exhaust control for transient heat load management in compact fusion reactors}
	
\author*[1,2]{\fnm{B.} \sur{Kool}}\email{b.kool@differ.nl}
\author[3]{\fnm{K.} \sur{Verhaegh}}\email{kevin.verhaegh@ukaea.uk}
\author[1,2]{\fnm{G.L. Derks.}}
\author[1,2]{\fnm{T.A. Wijkamp.}}
\author[3,4]{\fnm{N. Lonigro}}
\author[3,6]{\fnm{R. Doyle}}
\author[3]{\fnm{G. McArdle}}
\author[3]{\fnm{C. Vincent}}
\author[5]{\fnm{J. Lovell}}
\author[5]{\fnm{F. Federici}}
\author[3]{\fnm{S.S. Henderson}}
\author[3]{\fnm{R.T. Osawa}}
\author[7]{\fnm{D. Brida}}
\author[8]{\fnm{H. Reimerdes}}
\author[1]{\fnm{M. van Berkel}}
\author[{ }]{\fnm{The EUROfusion tokamak exploitation team$^\ast$}}
\author[{ }]{\fnm{and the MAST-U team$^\dagger$}}

\affil[1]{\orgname{DIFFER - Dutch Institute for Fundamental Energy Research}, \orgaddress{ \city{Eindhoven}, \country{the Netherlands}}}
\affil[2]{\orgname{Eindhoven University of Technology}, \orgaddress{\city{Eindhoven},\country{the Netherlands}}}
\affil[3]{\orgname{UKAEA-CCFE Culham Science Centre}, \orgaddress{\city{Culham}, \country{United Kingdom of Great Britain and Northern Ireland}}}
\affil[4]{\orgname{York Plasma Institute}, \orgaddress{\city{York},\country{United Kingdom of Great Britain and Northern Ireland}}}
\affil[5]{\orgname{Oak Ridge National Laboratory},  \orgaddress{\city{Oak Ridge}, \country{USA}}}
\affil[6]{\orgname{National Centre for Plasma Science and Technology},  \orgaddress{\city{Dublin}, \country{Ireland}}}
\affil[7]{\orgname{Max–Planck-Institut fur Plasmaphysik},  \orgaddress{\city{Garching}, \country{Germany}}}
\affil[8]{\orgname{EPFL-SPC École Polytechnique Fédérale de Lausanne},  \orgaddress{\city{Lausanne}, \country{Switzerland}}}
\affil[{ }]{\orgname{$^\ast$See author list of E. Joffrin et al. Nuclear Fusion 2024, doi: 10.1088/1741-4326/ad2be4}}
\affil[{ }]{\orgname{$^\dagger$See author list of J. Harrison et al. Nuclear Fusion 2019, doi: 10.1088/1741-4326/ab121c}}


\abstract{
 Nuclear fusion could offer clean, abundant energy. However, managing the immense power exhausted from the core fusion plasma towards the divertor remains a major challenge. This is compounded in emerging compact reactor designs which promise more cost-effective pathways towards commercial fusion energy. Alternative divertor configurations (ADCs) are a potential solution to this challenge. In this work, we demonstrate exhaust control in ADCs for the first time, on MAST-U. We employ a novel diagnostic strategy for the neutral gas buffer which shields the target. Our work shows that ADCs tackle key risks and uncertainties in realising fusion energy: 1) an enlarged operating window which 2) improves exhaust control through the absorption of transients which can remove the neutral shield and damage the divertor, 3) isolation of each divertor from other reactor regions, enabling combined control. This showcases real-world benefits of alternative divertors for effective heat load management and control in reactors.
}

\maketitle

\begin{figure*}[t]
	\makebox[1.1\textwidth][c]{
		\centering
		\hspace{-1.8cm}
		\includegraphics[width=\textwidth]{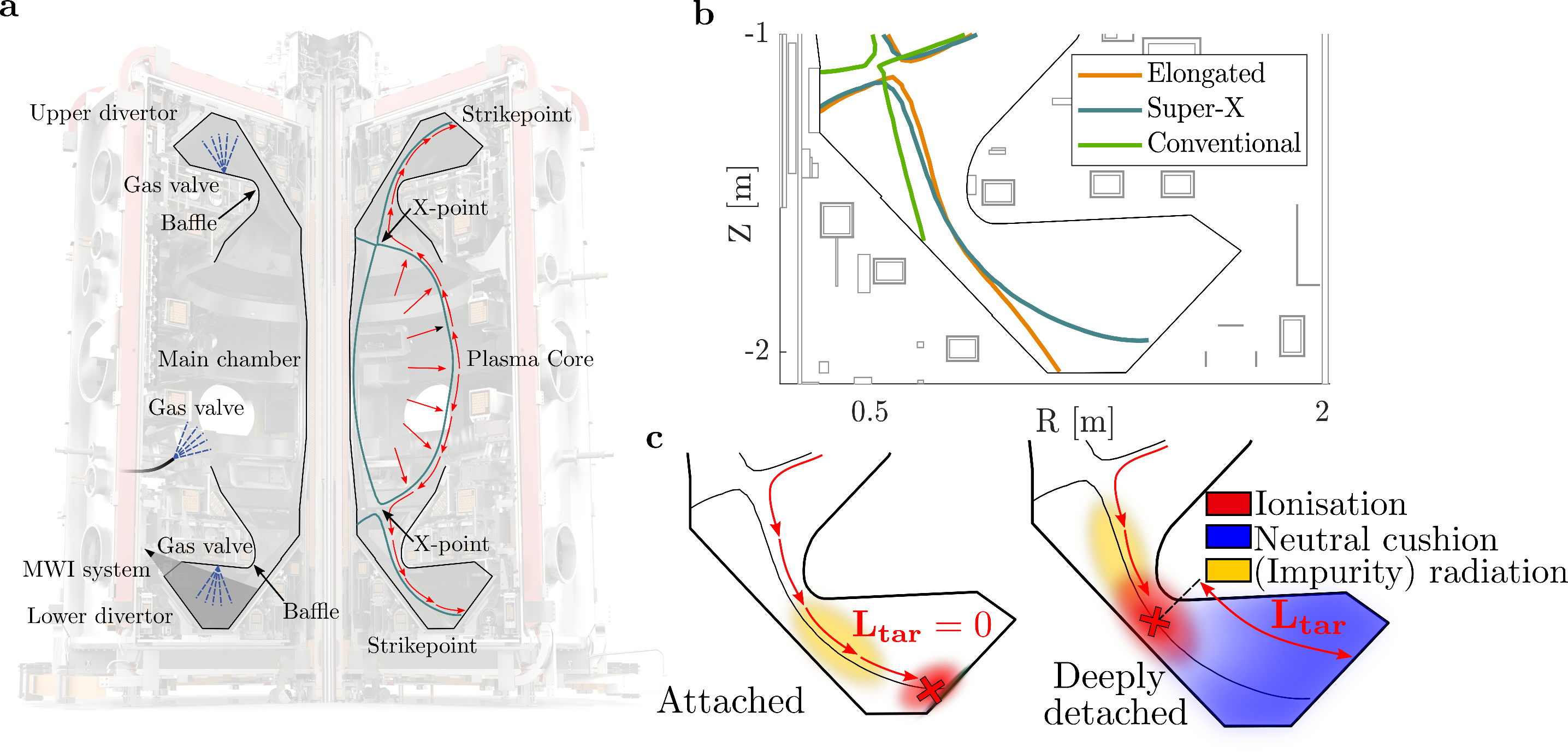}}
	\caption{\textbf{The MAST-U tokamak, its divertor configurations and detachment}. \textbf{a} MAST-U vessel\cite{MAST-Uimagegallery} with the gas valve and MWI\cite{Feng2021Development11channelmultiwavelengthimagingdiagnosticdivertorplasmasMASTUpgrade} locations indicated, the transport of energetic core particles towards the outer divertor targets is indicated by the red arrows. \textbf{b} Magnetic equilibria of the Super-X, Elongated, and Conventional divertor configurations.   \textbf{c} Schematic of an attached divertor (left) with volumetric impurity radiation and an ionisation region located at the strikepoint ($\mathrm{L}_{\mathrm{tar}}=0$), a detached  ($\mathrm{L}_{\mathrm{tar}}>0$) divertor (right) absorbs the incoming heat flux by forming a neutral cushion, massively reducing the strikepoint heat flux.\label{fig:overview}}
\end{figure*}

Nuclear fusion has the potential to provide virtually limitless, inherently safe, and clean energy \cite{Cowley2016questfusionpower}. The leading configuration, the tokamak \cite{Wesson2004Tokamaks}, magnetically confines particles in a torus-shaped device (Fig. \ref{fig:overview}a). However, the core power and plasma particles eventually propagate outside the magnetically confined region and are compressed into a narrow region of open magnetic field lines, resulting in an immense heat flux \cite{Zohm2021EUstrategysolvingDEMOexhaustproblem} akin to a welding torch. To prevent this heat flux from reaching the main chamber wall (Fig. \ref{fig:overview}a), it is diverted to a dedicated region (the divertor - Fig. \ref{fig:overview}a) using coils to create a magnetic 'null'. A key issue for reactor-scale devices is that the resulting divertor heat load far exceeds material limits if not mitigated \cite{Zohm2021EUstrategysolvingDEMOexhaustproblem}. 

This paper demonstrates for the first time the viability and successful control of power exhaust in two different alternative divertor configurations, solidifying them as a solution to this critical issue towards fusion energy, an absolute necessity for compact reactor development.  Compact high magnetic field reactors such as SPARC\cite{Creely2020OverviewSPARCtokamak}, ARC\cite{Sorbom2015ARCcompacthighfieldfusionnuclearsciencefacilitydemonstrationpowerplantdemountablemagnets}, and STEP\cite{Wilson2021STEPpathwayfusioncommercialization,Meyer2023Proceedings29thIAEAFusionEnergyConference,Muldrew2024ConceptualdesignworkflowSTEPPrototypePowerplant}, promise a more cost-effective and faster route to commercial fusion energy \cite{PengSPHERICALTORUSCOMPACTFUSIONLOWHELD}. However, the decreased size of these compact reactors results in a significantly increased heat and particle load per area, exacerbating the heat exhaust challenge\cite{Hudoba2023Divertoroptimisationpowerhandlingsphericaltokamakreactors,Kuang2020DivertorheatfluxchallengemitigationSPARC} which is already daunting for large reactors such as ITER\cite{Pitts2019PhysicsbasisfirstITERtungstendivertor} and DEMO\cite{Zohm2021EUstrategysolvingDEMOexhaustproblem}. Therefore, all these devices rely on, and necessitate, the development of Alternative Divertor Configurations (ADCs)\cite{Havlickova2015effectSuperXdivertorMASTUpgradeimpurityradiationmodelledSOLPS,Theiler2017ResultsrecentdetachmentexperimentsalternativedivertorconfigurationsTCV} to manage the substantial heat and particle load effectively. As these loads are not only static but also change dynamically, the control of transients is critical. To test both the passive mitigation and active transient suppression, the Mega Ampere Spherical Tokamak in the UK was recently upgraded (MAST-U) \cite{Morris2018MASTUpgradeDivertorFacilityTestBedNovelDivertorSolutions,Fishpool2013MASTupgradedivertorfacilityassessingperformancelongleggeddivertors,Harrison2019OverviewnewMASTphysicsanticipationfirstresultsMASTUpgrade} (extended data   Fig. \ref{edfig:mast-u}) to facilitate long-legged alternative divertor configurations such as the Elongated (ED) and Super-X (SXD) divertor. 

In a divertor, the plasma heat flux can be reduced by orders of magnitude by converting the hot plasma into photons and neutrals, spreading the heat flux over a larger area. This is achieved by injecting impurity and/or hydrogen gasses, which first dissipates the majority of the power ($\sim75 \%$) \cite{Verhaegh2023roleplasmaatommoleculeinteractionspowerparticlebalancedetachmentMASTUpgradeSuperXdivertor} in a \emph{radiative regime} upstream of the ion source, which remains 'attached' at the target ($\mathrm{T}_\mathrm{e} > 3-5$ eV). By 'detaching' the ionising plasma from the strike point with further gas injection, a cushion of neutral atoms and molecules is built downstream ($\mathrm{T}_\mathrm{e} < 3-5$ eV): \emph{detached regime}, enabling further power removal that can ultimately enable the required order-of-magnitude heat flux reduction.

Insufficient gas injection leads to insufficient dissipation and as such can destroy the divertor, while excessive gas injection can drive the cold, detached region into the core, potentially terminating the fusion plasma and damaging the reactor \cite{RavensbergenControl2021,theilerResultsRecentDetachment2017}. The exhaust challenge is exacerbated by transients originating from plasma core instabilities \cite{Wesson2004Tokamaks}, fuelling effects \cite{Wiesen2017ControlparticlepowerexhaustpelletfuelledITERDTscenariosemployingintegratedmodels}, and power sharing imbalances between various targets \cite{DeTemmerman2011HeatloadasymmetriesMAST,Osawa2023SOLPSITERanalysisproposedSTEPdoublenullgeometryimpactdegreedisconnectionpowersharing}. Such transients can cause cyclic power loading that leads to divertor tile cracking \cite{Morgan2021CombinedhighfluencehighcyclenumbertransientloadingITERlikemonoblocksMagnumPSI,Yu2015effecttransienttemporalpulseshapesurfacetemperaturetungstendamage,Li2020Fracturebehaviortungstenbasedcompositesexposedsteadystatetransienthydrogenplasma,Li2021PowerdepositionbehaviorhighdensitytransienthydrogenplasmatungstenMagnumPSI} and a loss of stable detachment. Active power exhaust control is thus imperative for reactor-scale devices to achieve the 'right' balance \cite{RavensbergenControl2021,Lipschultz2016Sensitivitydetachmentextentmagneticconfigurationexternalparameters,Zohm2021EUstrategysolvingDEMOexhaustproblem,Biel2022DevelopmentconceptbasisDEMOdiagnosticcontrolsystem} and requires knowledge of both the \emph{radiative} and \emph{detached} regimes. 

Recent progress in power exhaust control has been achieved using conventional, single null divertors, using diagnostics such as impurity emission \cite{Koenders2022SystematicextractioncontrolorientedmodelperturbativeexperimentsSOLPSITERemissionfrontcontrolTCV,RavensbergenControl2021}, tile temperature \cite{Brunner2016FeedbacksystemdivertorimpurityseedingbasedrealtimemeasurementssurfaceheatfluxAlcatorCModtokamaka}, plasma temperature \cite{Kolemen2015Heatfluxmanagementadvancedmagneticdivertorconfigurationsdivertordetachment}, radiation \cite{Kallenbach2015PartialdetachmenthighpowerdischargesASDEXUpgrade,Eldon2019AdvancesradiatedpowercontrolDIIID,Bernert2021XpointradiationitscontrolELMsuppressedradiatingregimeASDEXUpgradetokamak,Xu2020DivertorimpurityseedingnewfeedbackcontrolschememaintaininggoodcoreconfinementgrassyELMHmoderegimetungstenmonoblockdivertorEAST}, ion target flux \cite{Guillemaut2017RealtimecontroldivertordetachmentHmodeimpurityseedingusingLangmuirprobefeedbackJETITERlikewall}, and tile current \cite{Kallenbach2012OptimizedtokamakpowerexhaustdoubleradiativefeedbackASDEXUpgradea}. However, active control can only mitigate transients which are slow enough for gas actuators to respond, faster transients must be absorbed passively and this capability is only limited. It therefore remains uncertain whether the power exhaust challenge can be effectively managed using conventional divertors  \cite{Reimerdes2020AssessmentalternativedivertorconfigurationsexhaustsolutionDEMO,theilerResultsRecentDetachment2017}, posing a major risk for fusion power reactors.

Several alternative exhaust solutions are being explored as a risk mitigation strategy. This includes liquid metal divertor targets \cite{Sizyuk2022Liquidlithiumdivertormaterialmitigateseveredamagenearbycomponentsplasmatransients,vanEden2017Oscillatoryvapourshieldingliquidmetalwallsnuclearfusiondevices}, highly radiative plasma regimes \cite{Bernert2023XPointradiatingregimeASDEXUpgradeTCV,Field2017DynamicsstabilitydivertordetachmentHmodeplasmasJET,Gloggler2019CharacterisationhighlyradiatingneonseededplasmasJETILW}, and ADCs which rely on plasma shaping to improve power exhaust, detachment access, and its control \cite{Havlickova2015effectSuperXdivertorMASTUpgradeimpurityradiationmodelledSOLPS,Theiler2017ResultsrecentdetachmentexperimentsalternativedivertorconfigurationsTCV}. 

One of the most prominent ADCs is the Super-X Divertor (SXD) (Fig. \ref{fig:overview}a) featuring an increased strike point radius, increasing the target area. The MAST-U tokamak uniquely enables integrating ADCs with physical structures (baffles\cite{Reimerdes2021InitialTCVoperationbaffleddivertor,Havlickova2015SOLPSanalysisMASTUdivertoreffectheatingpowerpumpingaccessdetachmentSuperxconfiguration}) to prevent the escape of divertor neutrals towards the core in a symmetric up/down double-null divertor for improved power handling and H-mode access \cite{Osawa2023SOLPSITERanalysisproposedSTEPdoublenullgeometryimpactdegreedisconnectionpowersharing,Wigram2019PerformanceassessmentlongleggedtightlybaffleddivertorgeometriesARCreactorconcept,Kuang2020DivertorheatfluxchallengemitigationSPARC,Reimerdes2020AssessmentalternativedivertorconfigurationsexhaustsolutionDEMO}.  Under steady-state conditions, massively reduced target heat and particle fluxes with an improved access to detachment has been observed \cite{Verhaegh2024InvestigationsofatomicmolecularprocessesofNBI-heateddischargesintheMASTUpgradeSuper-Xdivertorwithimplicationsforreactors,Verhaegh2024Long-leggedDivertorsandNeutralBafflingasaSolutiontotheTokamakPowerExhaustChallenge,Verhaegh2023roleplasmaatommoleculeinteractionspowerparticlebalancedetachmentMASTUpgradeSuperXdivertor}, consistent with reduced model predictions \cite{Lipschultz2016Sensitivitydetachmentextentmagneticconfigurationexternalparameters} and simulations \cite{Havlickova2015SOLPSanalysisMASTUdivertoreffectheatingpowerpumpingaccessdetachmentSuperxconfiguration,Havlickova2015effectSuperXdivertorMASTUpgradeimpurityradiationmodelledSOLPS}.

In this work, we demonstrate exhaust control in ADCs for the first time, highlighting the intrinsic capability of the SXD to passively absorb transients and therefore present a solution to the heat exhaust problem. Our results also show that the upper and lower divertors are largely decoupled, suggesting that both divertors can be independently controlled, a requirement for double-null reactor designs \cite{Osawa2023SOLPSITERanalysisproposedSTEPdoublenullgeometryimpactdegreedisconnectionpowersharing}. Unlike previously employed indirect methods \cite{Brunner2016FeedbacksystemdivertorimpurityseedingbasedrealtimemeasurementssurfaceheatfluxAlcatorCModtokamaka,Bernert2021XpointradiationitscontrolELMsuppressedradiatingregimeASDEXUpgradetokamak,RavensbergenControl2021,Kallenbach2012OptimizedtokamakpowerexhaustdoubleradiativefeedbackASDEXUpgradea,Guillemaut2017RealtimecontroldivertordetachmentHmodeimpurityseedingusingLangmuirprobefeedbackJETITERlikewall}, we achieve power exhaust control through active \emph{detachment control} through a novel, more direct, detection technique. Here, the $\mathrm{D}_2$ Fulcher emission intensity, originating from electronically excited molecules, serves as a proxy for the ionising region \cite{Verhaegh2023SpectroscopicinvestigationsdetachmentMASTUpgradeSuperXdivertor,Verhaegh2023roleplasmaatommoleculeinteractionspowerparticlebalancedetachmentMASTUpgradeSuperXdivertor}. 

This pioneering experimental work is a major step towards addressing one of the key challenges for realising fusion energy. The implications of the ADC benefits for power exhaust control for reactor designs are discussed and the scalability of our detachment sensor strategy to reactors are demonstrated on a scientific, conceptual level focusing on the STEP reactor. 

\section*{Divertor dynamics of strongly baffled Alternative Divertor Configurations}\label{sec:Dynamics}
In our experimental results, we observe major advantages of ADCs for solving the fusion exhaust challenge:  (1) an enlarged operating window which facilitates the absorption of transients and therefore (2) improves detachment control through a resilience to re-attachment and radiative collapse. The tight baffling isolates each divertor from other reactor regions which (3) enables combined control.

First, the \emph{divertor dynamics} are studied to investigate the capability of handling transients using a systematic system identification approach that has been applied previously to \emph{conventional divertors} in recent research\cite{RavensbergenControl2021,Koenders2022SystematicextractioncontrolorientedmodelperturbativeexperimentsSOLPSITERemissionfrontcontrolTCV,Koenders2023ModelbasedimpurityemissionfrontcontrolusingdeuteriumfuelingnitrogenseedingTCV}. In these experiments, the divertor response to deuterium gas inflow perturbations is measured to identify the dynamics. Using system identification and novel ionisation front tracking techniques, we perform a pioneering study of the \emph{detachment front} dynamics in \emph{alternative divertor configurations} in MAST-U. 

The dynamics of divertor detachment are studied in three MAST-U double null divertor scenarios: the Conventional Divertor (CD), the Elongated (ED) and Super-X (SXD) alternative divertor configurations (Fig. \ref{fig:overview}b). We study the dynamical response of divertor detachment to perturbations in the main chamber $\mathrm{D}_2$ flow rate request in a low-confinement mode (L-mode), NBI-heated scenario (see methods).

The detachment state is observed by tracking the ionisation front using the lower divertor (Fig. \ref{fig:overview}a,b) Multi-Wavelength-Imaging (MWI) \cite{Feng2021Development11channelmultiwavelengthimagingdiagnosticdivertorplasmasMASTUpgrade} camera measurements of the  $\mathrm{D}_2$ Fulcher emission as an ionisation proxy \cite{Verhaegh2023SpectroscopicinvestigationsdetachmentMASTUpgradeSuperXdivertor,Verhaegh2023roleplasmaatommoleculeinteractionspowerparticlebalancedetachmentMASTUpgradeSuperXdivertor}. These images are inverted\cite{Wijkamp2023CharacterisationdetachmentMASTUSuperXdivertorusingmultiwavelengthimaging2Datomicmolecularemissionprocesses} to obtain a 2D emissivity distribution, enabling the ionisation front position to be tracked\cite{Wijkamp2023CharacterisationdetachmentMASTUSuperXdivertorusingmultiwavelengthimaging2Datomicmolecularemissionprocesses}. The (poloidal) distance between the target and the ionisation front  along the divertor leg is quantified: $\mathrm{L}_{\mathrm{tar}}$ (Fig. \ref{fig:overview}b) ($\mathrm{L}_{\mathrm{tar}}>0$ during detachment, see methods).

\begin{figure}
	\makebox[0.5\textwidth][c]{
		\centering
		\includegraphics[width=0.5\textwidth]{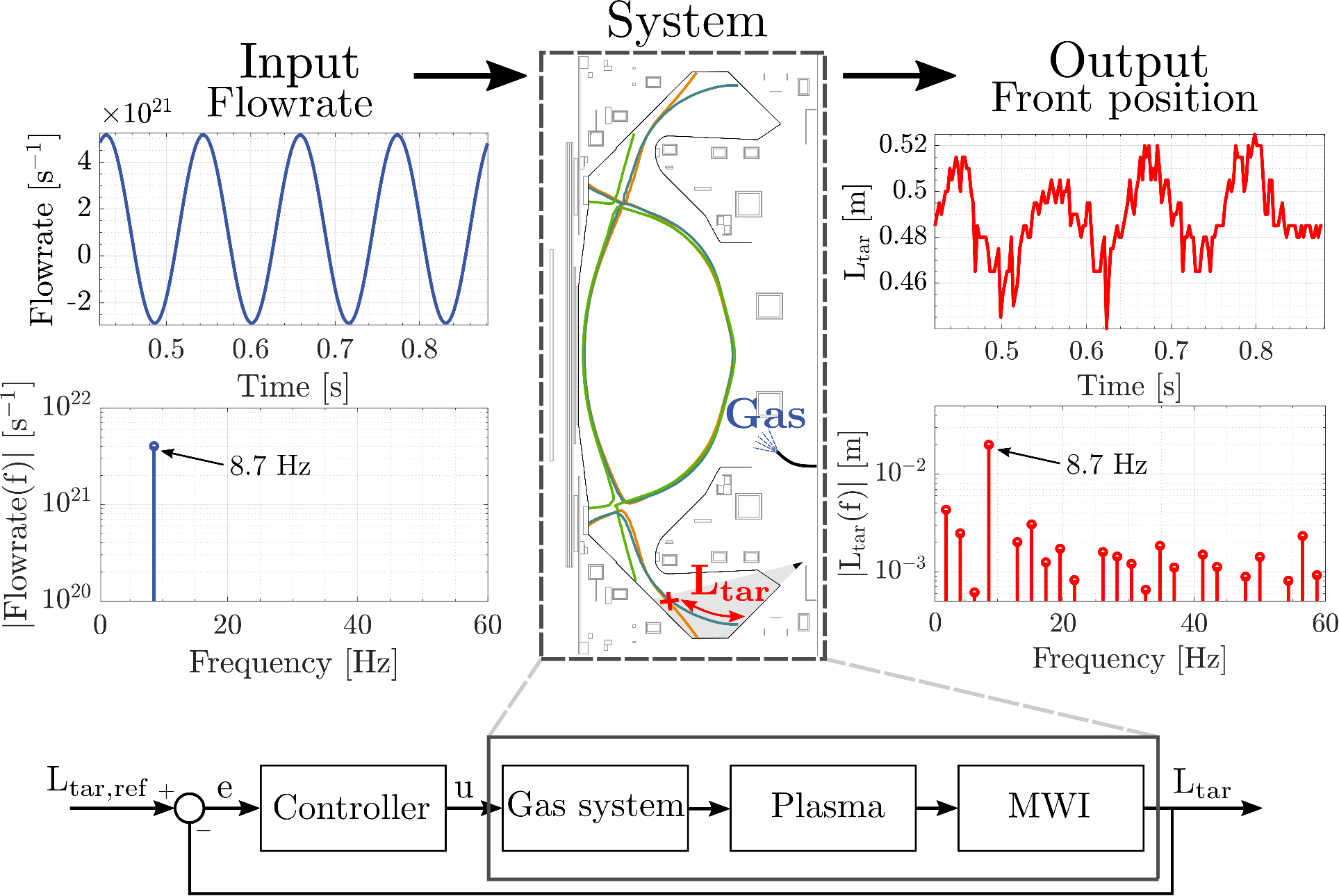}}
	\caption{\textbf{System identification and the feedback control loop}.  (top) Time domain and frequency domain response of the $\mathrm{D}_2$ Fulcher band front position $\mathrm{L}_{\mathrm{tar}}$ to a perturbation of the requested main chamber $\mathrm{D}_2$ gas valve flow rate in Super-X divertor geometry, shot \#47116. The input frequency component is clearly observed in the system output, see also extended data   Fig. \ref{edfig:frontresponse}. (bottom) Feedback control loop showing the reference front position $\mathrm{L}_{\mathrm{tar,ref}}$, the error $\mathrm{e}$ with respect to the measured front position $\mathrm{L}_{\mathrm{tar}}$, and the corresponding requested flow rate $\mathrm{u}$ by the controller. The to-be-controlled system includes the gas system, plasma, and MWI sensor dynamics.  }
	\label{fig:lines} 
\end{figure}

By analysing the frequency response of $\mathrm{L}_{\mathrm{tar}}$, information on the detachment front dynamics are retrieved. A typical time and frequency domain response is indicated in Fig. \ref{fig:lines}. The perturbation signals are especially designed to optimise the signal-to-noise ratio  within experimental constraints (see methods). The considered system includes the MWI camera, plasma response, and gas system dynamics and is conceptually illustrated in Fig. \ref{fig:lines}. Since only the \emph{requested} gas valve flow rate can be perturbed, the gas system dynamics cannot be separated from the detachment dynamics \cite{Derks2024DevelopmentrealtimedensityfeedbackcontrolMASTULmode} (see methods). A first result from these studies is that the response on non-excited frequencies does not exceed the noise floor, indicating that the dynamics for both Elongated (Fig. \ref{fig:lines}) and Super-X configuration (extended data   Fig.\ref{edfig:frontresponse}) are predominantly linear \cite{Schoukens2009}. This crucial finding implies that standard, linear control techniques are also suitable for these alternative divertor configurations, akin to conventional configurations\cite{RavensbergenControl2021,Koenders2023ModelbasedimpurityemissionfrontcontrolusingdeuteriumfuelingnitrogenseedingTCV}.

\begin{figure}
	\centering
	\includegraphics[width=0.5\textwidth]{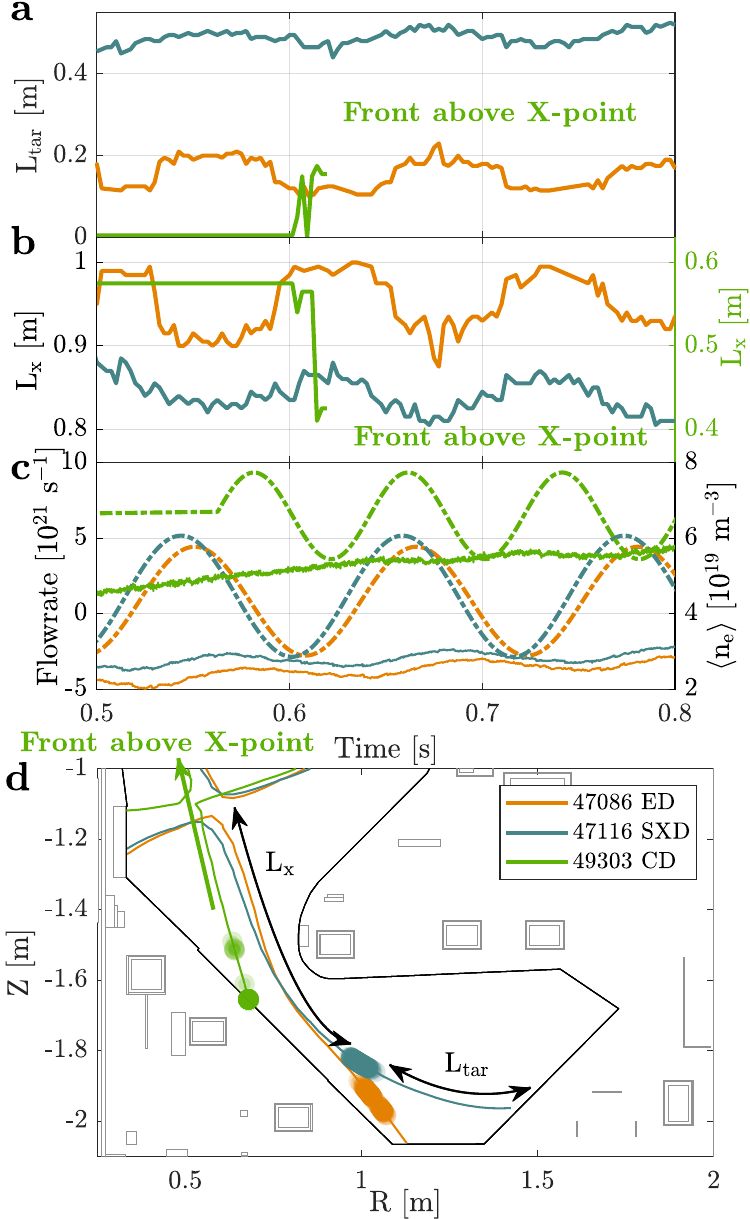}
	\caption{\textbf{Comparison of Super-X, Elongated, and Conventional divertor front movement}. \textbf{a} Divertor detachment state expressed as distance of the Fulcher band front position to target along the divertor leg $\mathrm{L}_{\mathrm{tar}}$. \textbf{b} $\mathrm{D}_2$ Fulcher front position as Poloidal distance between the $\mathrm{D}_2$ Fulcher emission front and the X-point $\mathrm{L}_{\mathrm{x}}$. \textbf{c} Requested main chamber gas flow rate (left, -) and line-averaged core electron density (right, \texttt{-{}-}). \textbf{d} Two-dimensional view of the magnetic configurations and front position movement within the time frame. }
	\label{fig:SXDvsECD}
\end{figure}

\begin{figure*}
	\hspace{-2cm}
	\makebox[1.1\textwidth][c]{
		\centering
		\includegraphics[width=\textwidth]{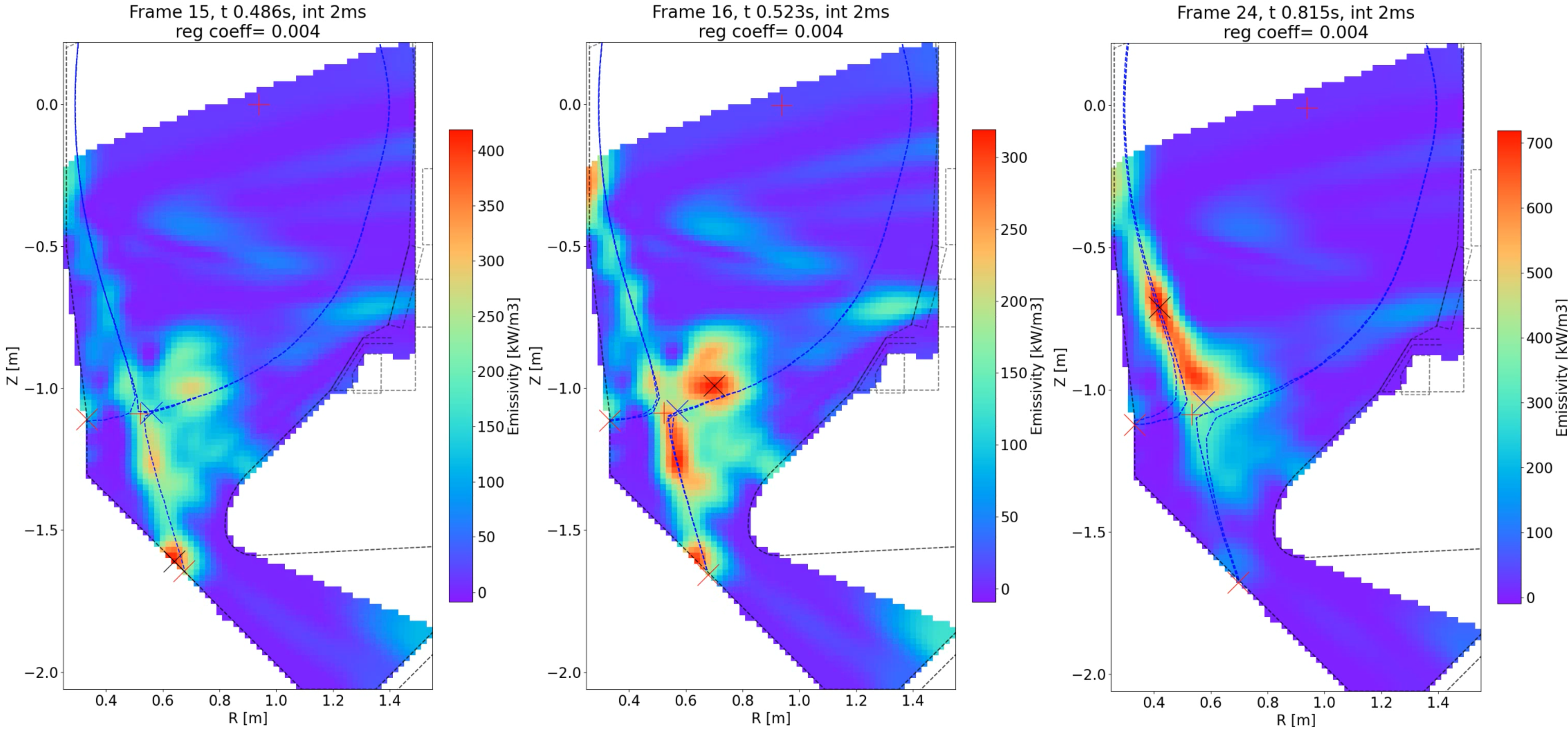}}
	\renewcommand\figurename{Extended data fig}
	\caption{\textbf{\hspace{-0.2cm}$|$ Imaging bolometry \#49303 } Progression of radiation location as measured by the IRVB imaging bolometry system\cite{Federici2023DesignimplementationprototypeinfraredvideobolometerIRVBMASTUpgrade}, indicating a transition from mainly attached radiation near the strikepoint to volumetric radiation at the high-field-side.}
	\label{edfig:IRVB}
\end{figure*}

Comparing the dynamical response of the three divertor configurations shows that the Elongated and Super-X divertors have a much greater capacity to passively absorb transients and feature a reduced detachment onset. In contrast, the conventional divertor requires much higher baseline flow rates ($5\cdot10^{21}$ $\mathrm{D}_2$/s) and line-averaged core densities ($5\cdot10^{19}$ $\mathrm{m}^{-3}$ vs $3\cdot10^{19}$ $\mathrm{m}^{-3}$) to detach, whereas the ED and SXD are already detached at the lowest flow rate request to maintain stable core conditions ($10^{21}$ $\mathrm{D}_2$/s) (Fig. \ref{fig:SXDvsECD}a, showing $\mathrm{L}_{\mathrm{tar}}>0$ for the ED and SXD). When the ionisation front starts to detach from the target in the conventional divertor ($\mathrm{L}_{\mathrm{tar}}>0$), it quickly moves outside the operational window, towards the X-point (poloidal distance between the ionisation front and the X-point, $\mathrm{L}_{\mathrm{x}}\approx0$), resembling a MARFE\cite{Lipschultz1984Marfeedgeplasmaphenomenon} (see extended data Fig. \ref{edfig:IRVB} for bolometry measurements). This indicates a very narrow Conventional divertor detachment window and inherently, a much more pronounced response to fuelling perturbations. Additionally, when the fuelling is subsequently reduced, the MARFE remains. 

This is in stark contrast with both the Super-X and Elongated divertors, where the response to the perturbations is only minimal (limited $\mathrm{L}_{\mathrm{tar}}$ variation). The combination of this decreased sensitivity and increased divertor leg length entails that ADCs have a much larger capacity to passively absorb transients and therefore avoid re-attachment and a radiative core plasma collapse. These findings are consistent with previous steady-state observations \cite{Verhaegh2023roleplasmaatommoleculeinteractionspowerparticlebalancedetachmentMASTUpgradeSuperXdivertor,Verhaegh2023SpectroscopicinvestigationsdetachmentMASTUpgradeSuperXdivertor} of a reduced detachment onset and front sensitivity to steady-state density variations for ADCs. These pioneering results now illustrate for the first time that these benefits: a major reduction in detachment front sensitivity for alternative divertors, extend to a dynamic situation.

\begin{figure}
		\makebox[0.48\textwidth][c]{
		\centering
		\includegraphics[width=0.5\textwidth]{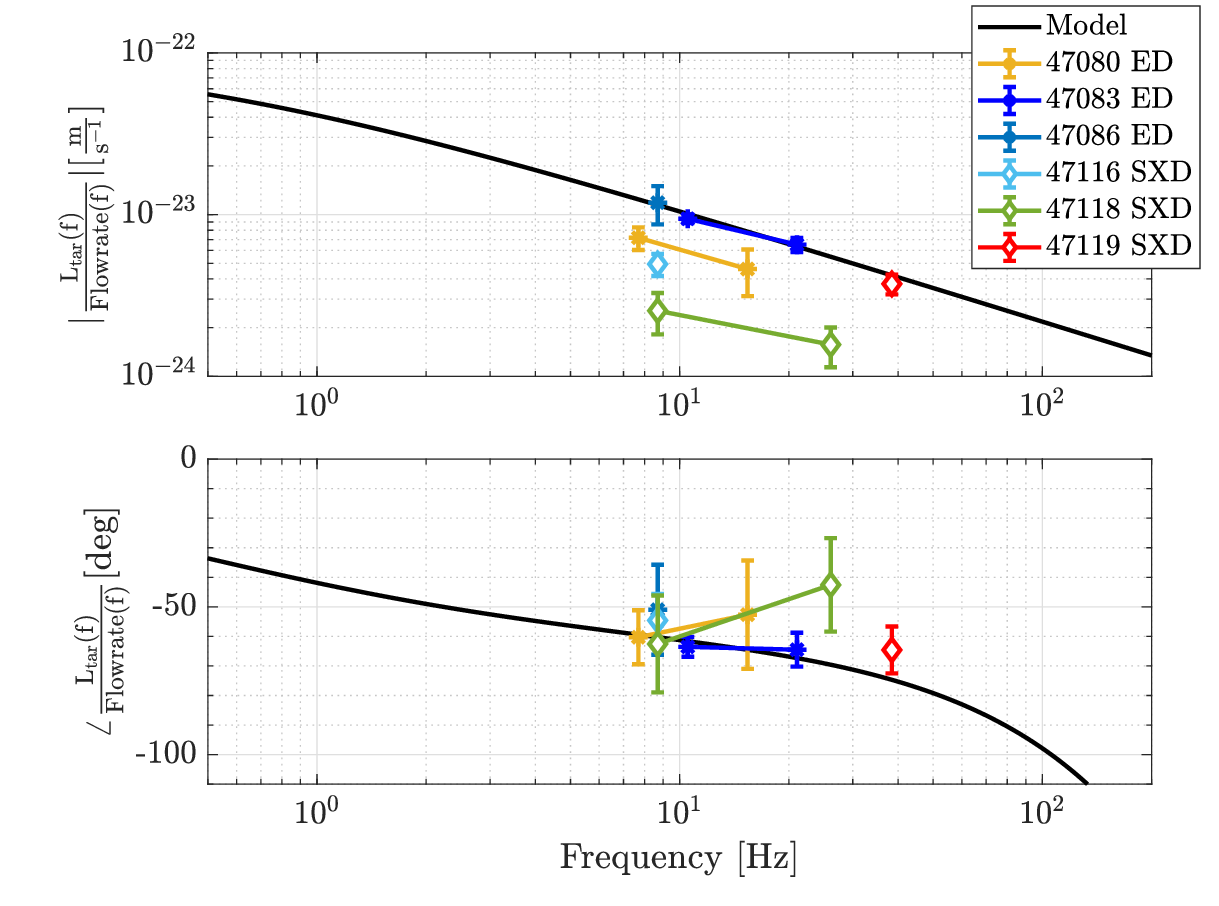}}
	\caption{\textbf{Identified frequency response}. Requested main chamber gas valve flow rate to Fulcher band front position $\mathrm{L}_{\mathrm{tar}}$ response in Elongated ($*$) and Super-X ($\lozenge$) configuration expressed as gain (top) and phase (bottom) ratio over frequency. Response and 2$\sigma$ error bars obtained using the LPM\cite{Schoukens2009,VanBerkel2020CorrectingnonperiodicbehaviourperturbativeexperimentsApplicationheatpulsepropagationmodulatedgaspuffexperiments}}
		\label{fig:systemID}
\end{figure}

As the ionisation front moved past the operational window in the Conventional divertor, only the dynamics for the Elongated and Super-X configuration can be identified. The linear dynamics of the Elongated ($*$) and Super-X ($\lozenge$) configurations around the operating point can be expressed as a frequency response function (FRF) by analysing the input-output ratio in frequency domain (Fig. \ref{fig:systemID}). We use the local-polynomial method\cite{Schoukens2009} (LPM) to correct for transient effects and to estimate 2$\sigma$ error bars \cite{VanBerkel2020CorrectingnonperiodicbehaviourperturbativeexperimentsApplicationheatpulsepropagationmodulatedgaspuffexperiments} (see methods). Across all measurements, we observe a phase delay between 40 and 70 degrees, indicative of a fractional order transfer. This suggests that the underlying physical processes can be modelled as a fractional differential equation\cite{SchoolofEngineeringandPhysicsTheUniversityoftheSouthPacificLaucalacampusFiji.2019FractionalOrderSystemModelingitsApplications}, inline with conventional divertor results\cite{RavensbergenControl2021,Koenders2022SystematicextractioncontrolorientedmodelperturbativeexperimentsSOLPSITERemissionfrontcontrolTCV}. As the origin of these dynamic is not yet understood, further investigation across multiple devices is foreseen. When comparing the Super-X and Elongated divertor configurations, we observe a similar phase response, however, the Super-X magnitude appears smaller compared to the Elongated divertor, as observed in Fig. \ref{fig:SXDvsECD}. This suggests a potential additional reduction in front sensitivity for the Super-X which warrants further study.

This marks the first, crucial, experimental confirmation of the benefits of ADCs in passively handling \emph{dynamic} transients, previously forecasted for steady-state conditions \cite{Havlickova2015SOLPSanalysisMASTUdivertoreffectheatingpowerpumpingaccessdetachmentSuperxconfiguration,Havlickova2015effectSuperXdivertorMASTUpgradeimpurityradiationmodelledSOLPS,Lipschultz2016Sensitivitydetachmentextentmagneticconfigurationexternalparameters}. In addition to demonstrating the benefits of ADCs for solving the exhaust challenge in fusion power reactors, our results also enable feedback exhaust control in MAST-U through the systematic \textit{design} of an exhaust controller.

\section*{Feedback control}\label{sec:Feedback}
We will now demonstrate that ADCs are compatible with detachment control, a requirement for their application in fusion power reactors\cite{RavensbergenControl2021,Lipschultz2016Sensitivitydetachmentextentmagneticconfigurationexternalparameters,Zohm2021EUstrategysolvingDEMOexhaustproblem,Biel2022DevelopmentconceptbasisDEMOdiagnosticcontrolsystem}.

By matching a dynamic model to the observed system dynamics (Fig. \ref{fig:systemID}), detachment control is enabled without having to explicitly capture the intricacies of the underlying physical processes. A simple fractional order transfer function $\mathrm{G}(j\omega)$ is used for the dynamical model with gain $\mathrm{K}= 10^{-22}$ [-], time constant $\tau=0.5$ [s], time delay $\tau_{\mathrm{d}}= 10^{-3}$ [s] and frequency $\omega$ [rad/s]. 
\begin{equation}
	\hfil	\mathrm{G}(j\omega)=\frac{\mathrm{K}}{\mathrm{\tau} (j\omega)^{0.7}+1}e^{-\tau_\mathrm{d}\mathrm{j\omega}},
\end{equation}
Using $\mathrm{G}(j\omega)$, we \textit{design} a basic Proportional-Integral (PI) controller $\mathrm{C}(j\omega)$ through the loopshaping method \cite{Skogestad2005} as
\begin{equation}
	\hfil	\mathrm{C}(j\omega)=\mathrm{K}_{\mathrm{p}}+\frac{\mathrm{K}_{\mathrm{i}}}{j\omega},
\end{equation}
with proportional and integral gains as $\mathrm{K}_\mathrm{p}=5\cdot10^{22}$ and $\mathrm{K}_\mathrm{i}=3\cdot10^{24}$ respectively. The controller is targeted towards robustness for this proof of concept demonstration, evidenced by the closed-loop bandwidth of 9.5 Hz and 70\textdegree\ phase margin, indicating that future performance improvements are likely possible. 

Feedback control requires real-time inference of the ionisation front position. An inversion-less emission front tracking algorithm \cite{Ravensbergen2020Developmentrealtimealgorithmdetectiondivertordetachmentradiationfrontusingmultispectralimaging}, that directly operates on raw camera images, is adopted for the real-time implementation. Although yielding comparable results to the inverted technique, the coordinate transformation introduces additional noise \cite{Hommen2014Opticalboundaryreconstructionshapecontroltokamakplasmas,Ravensbergen2020Developmentrealtimealgorithmdetectiondivertordetachmentradiationfrontusingmultispectralimaging} (see methods).

\begin{figure*}
	\makebox[1.1\textwidth][c]{
		\hspace{-1.2cm}
		\centering
		\includegraphics[width=1.05\linewidth]{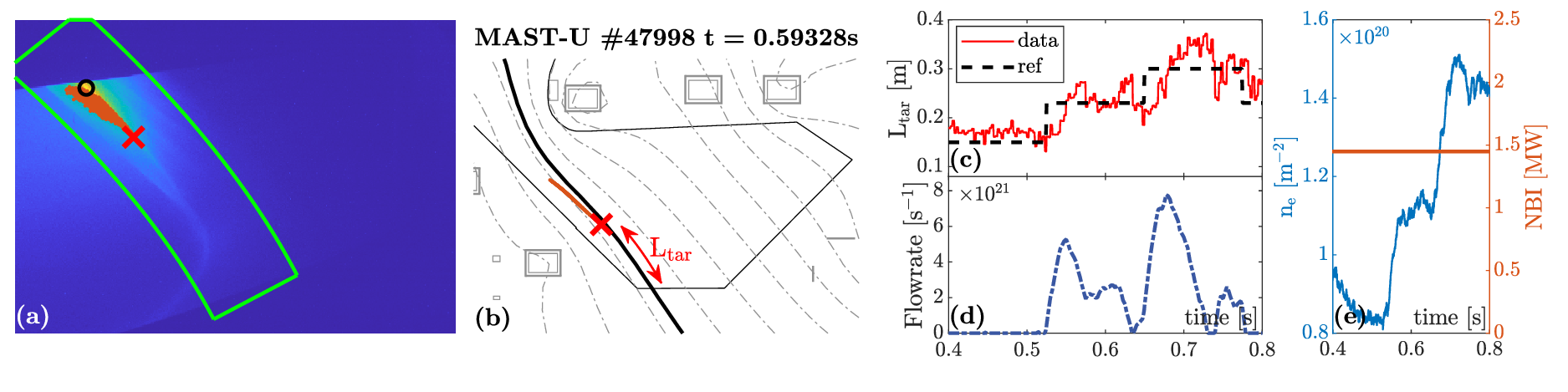}}
	
	\makebox[1.1\textwidth][c]{	
		\hspace{-1.2cm}
		\centering
	\includegraphics[width=1.05\linewidth]{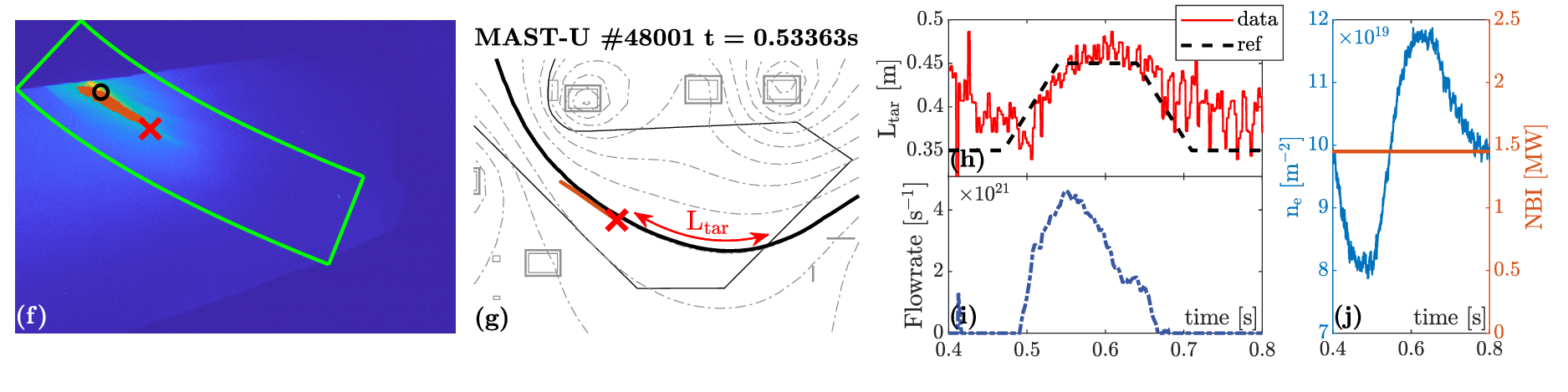}}
	\vspace{-0.4cm}
	\caption{\textbf{Feedback exhaust control in the lower divertor in MAST-U in Elongated (a-e) and Super-X (f-j) divertor configuration.} \textbf{a,e} Fulcher-band filtered MWI\cite{Feng2021Development11channelmultiwavelengthimagingdiagnosticdivertorplasmasMASTUpgrade} image with the detected emission front position defined as the 50\% extinction point (red cross) from the maximum intensity (the black circle). \textbf{b,g} Poloidal cross section showing the EFIT++ magnetic equilibrium and detected emission front position. \textbf{c,h} Time evolution of the poloidal distance-to-target $\mathrm{L}_{\mathrm{tar}}$ and its applied reference. \textbf{d,i} Time evolution of the flow rate request by the controller. \textbf{e,j} Time evolution of core line-integrated electron density and south-west neutral beam injector power.
		\label{fig:feedback}}
\end{figure*}

Detachment control is successfully achieved using the same controller and sensor for both Elongated and Super-X divertor configurations on MAST-U (Fig. \ref{fig:feedback}): in both scenarios, the controller adeptly follows the reference trajectory. This illustrates the robustness of our detachment control implementation: the same approach is successful for two distinct divertor configurations. In contrast, achieving detachment control in Conventional divertor configuration would be extremely challenging through the narrow operating range (Fig. \ref{fig:SXDvsECD}).

When the controller aims to move the front towards the target, the fuelling is successfully reduced. This, however, results in a slower movement of the ionisation front than requested (Fig. \ref{fig:feedback}f-j), as the divertor neutral pressure remains high given the lack of cryopumping capacity on MAST-U \cite{Morris2018MASTUpgradeDivertorFacilityTestBedNovelDivertorSolutions}. This illustrates the importance of adequate pumping for exhaust control, consistent with the impurity retention issues noted in TCV\cite{Koenders2023ModelbasedimpurityemissionfrontcontrolusingdeuteriumfuelingnitrogenseedingTCV}.

For the first time, we have demonstrated exhaust control in alternative divertor configurations. This proves the required compatibility with exhaust control, and therefore enables their application as a heat-exhaust solution for (compact) fusion power reactors. A key requirement for double-null power reactor designs is independent exhaust control of the lower/upper divertors to combat the expected asymmetric transients from divertor power sharing imbalances \cite{Osawa2023SOLPSITERanalysisproposedSTEPdoublenullgeometryimpactdegreedisconnectionpowersharing,Lennholm2024ControllingaNewPlasmaRegime}. Therefore, we investigate the coupling between upper and lower Super-X divertors in the next section.

\section*{Upper-lower divertor coupling}\label{sec:interaction}
We will demonstrate how strongly baffled ADCs possess key benefits for enabling independent control of multiple divertor regions through a reduction in the coupling between both divertors. Thus far, open divertor geometries have exhibited a clear coupling between upper and lower divertor states \cite{Fevrier2021a}. Here, we systematically investigate such coupling for the strongly baffled divertor chambers in MAST-U for the first time.

\begin{figure}
	\centering
	\includegraphics[width=0.5\textwidth]{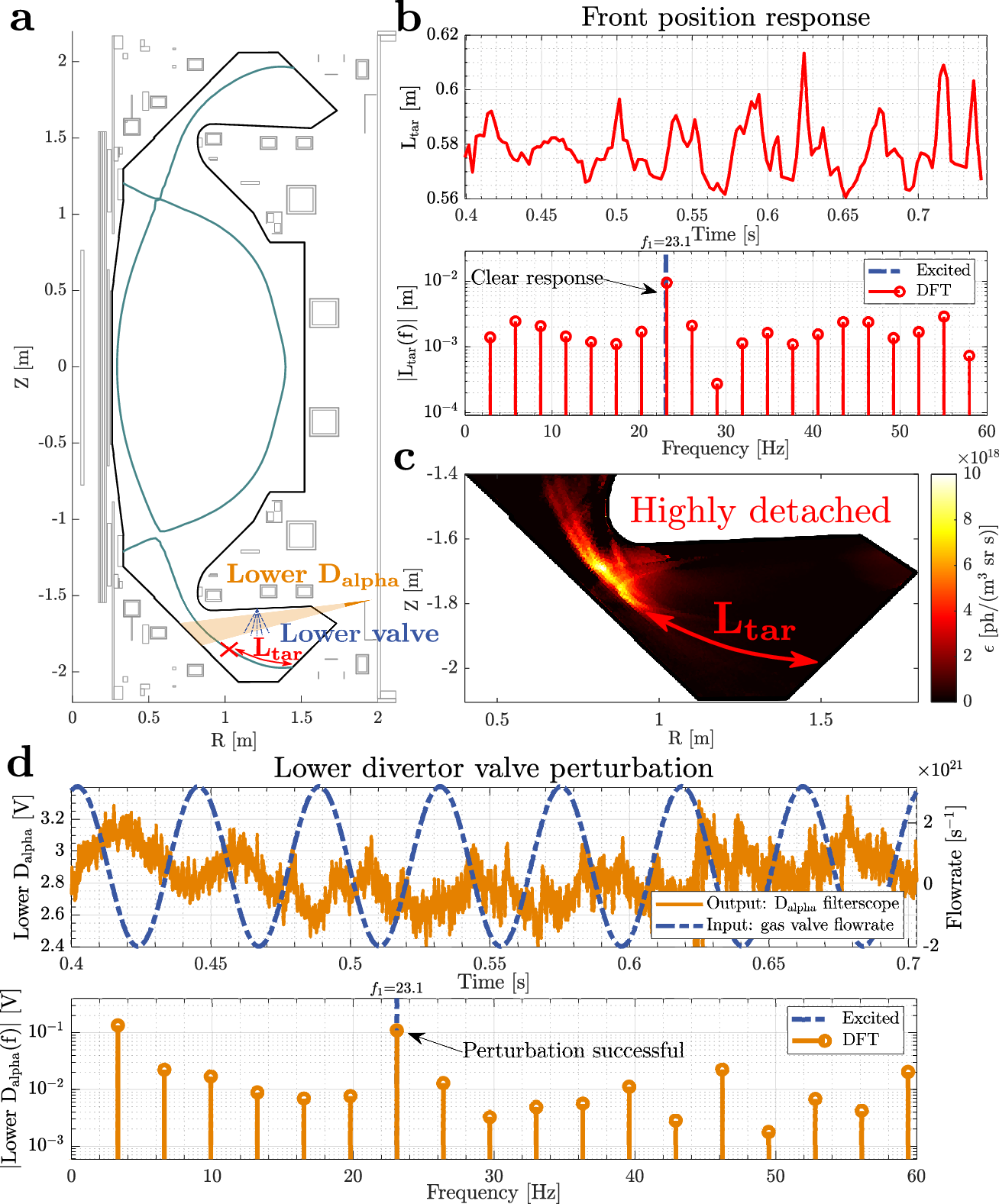} 
	\caption{\textbf{Lower divertor gas valve system identification in Super-X divertor configuration (\# 49297)} \textbf{a} Poloidal cross section showing the EFIT++ magnetic equilibrium  reconstruction ($t=0.55$ s) with a conceptual illustration of the lower divertor $\mathrm{D}_{\mathrm{alpha}}$ filterscope sightline, lower divertor valve location, and lower divertor Fulcher band front position $\mathrm{L}_{\mathrm{tar}}$.  \textbf{b} Time domain (top) and frequency domain (bottom) response of the de-trended lower divertor Fulcher band front position $\mathrm{L}_{\mathrm{tar}}$. \textbf{c} Combined MWI+XPI Fulcher band inversion ($t=0.55$ s), indicating highly detached conditions. \textbf{d} Time domain (top) and frequency domain (bottom) response of the de-trended lower divertor $\mathrm{D}_{\mathrm{alpha}}$ filterscope (-) to a perturbation of the requested lower deuterium divertor gas valve flow rate ({-}{-}).}	\label{fig:interaction lower}
\end{figure}

As opposed to perturbing core fuelling, individual system identification perturbations in the upper and lower divertors are performed. When perturbing \emph{only} the \emph{lower} divertor deuterium gas valve, clear responses are observed in both the \emph{lower} divertor ionisation front position $\mathrm{L}_{\mathrm{tar}}$ and \emph{lower} divertor $\mathrm{D}_{\mathrm{alpha}}$ filterscope emission (Fig. \ref{fig:interaction lower}), which is a key measure of plasma-neutral interaction. 

\begin{figure}
	\centering
	\includegraphics[width=0.5\textwidth]{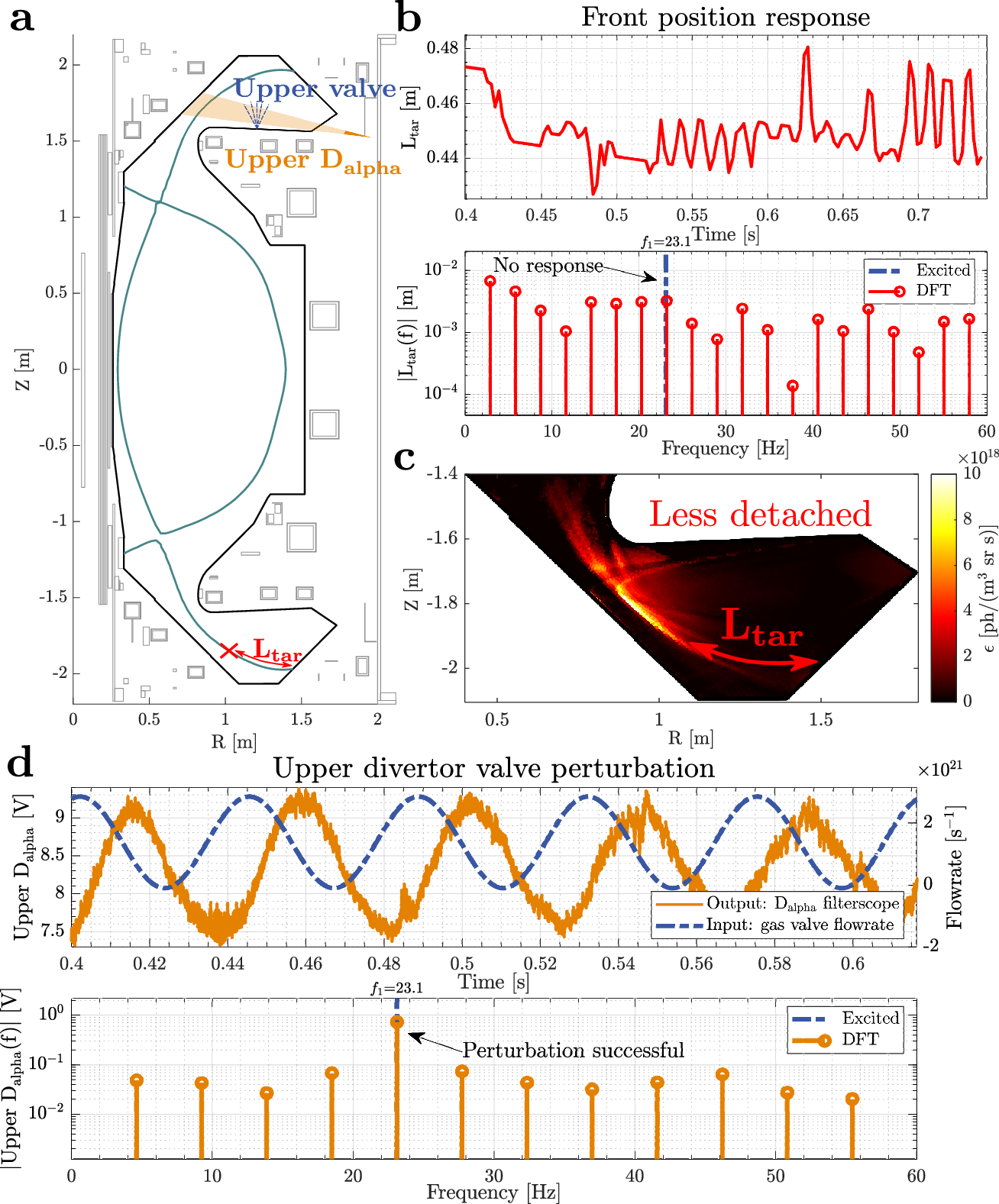}
			\caption{\textbf{Upper divertor gas valve system identification in Super-X divertor configuration (\# 49298)} \textbf{a} Poloidal cross section showing the EFIT++ magnetic equilibrium  reconstruction ($t=0.55$ s) with a conceptual illustration of the upper divertor $\mathrm{D}_{\mathrm{alpha}}$ filterscope sightline, upper divertor valve location, and lower divertor Fulcher band front position $\mathrm{L}_{\mathrm{tar}}$.  \textbf{b} Time domain (top) and frequency domain (bottom) response of the de-trended lower divertor Fulcher band front position $\mathrm{L}_{\mathrm{tar}}$. \textbf{c} Combined MWI+XPI Fulcher band inversion ($t=0.55$ s), indicating highly detached conditions. \textbf{d} Time domain (top) and frequency domain (bottom) response of the de-trended upper divertor $\mathrm{D}_{\mathrm{alpha}}$ filterscope (-) to a perturbation of the requested upper deuterium divertor gas valve flow rate ({-}{-}). }	\label{fig:interaction upper}
\end{figure}

Subsequently, \emph{only} the \emph{upper} divertor valve is perturbed (Fig. \ref{fig:interaction upper}). While we observe a clear response in the \emph{upper} $\mathrm{D}_{\mathrm{alpha}}$ filterscope intensity, \emph{no response} in the \emph{lower} divertor ionisation front position nor $\mathrm{D}_{\mathrm{alpha}}$ intensity is observed.  Additionally, the lower divertor is significantly less deeply detached than during the lower divertor perturbation (Fig. \ref{fig:interaction lower}). This disparity suggests a limited influence of the upper divertor perturbation on the lower divertor state, demonstrating a clear decoupling between the two divertors. This is a major result, confirming preliminary, static observations \cite{Verhaegh2023roleplasmaatommoleculeinteractionspowerparticlebalancedetachmentMASTUpgradeSuperXdivertor}.

Our results present a crucial finding for the deployment of ADCs in fusion power reactors as the required independent control of both divertor regions can likely be achieved. However, preliminary He seeding experiments have indicated that impurities injected in the lower divertor do spread rapidly throughout the vessel, potentially limiting the observed decoupling to hydrogen/neutral pressure only, prompting further study. The observed decoupling for deuterium fuelling is likely attributed to the strong neutral baffling of the divertor chamber as such decoupling is absent in open divertor geometries \cite{Fevrier2021a}. Our results, for the first time, prove that instabilities driven by interaction between the upper and lower divertor controllers are highly unlikely, allowing for independent control of the upper and lower Super-X divertors. 

\section*{Discussion \& Conclusion: Implications for power reactors}\label{sec:implications}

\begin{figure*}
		\centering
		\includegraphics[width=\linewidth]{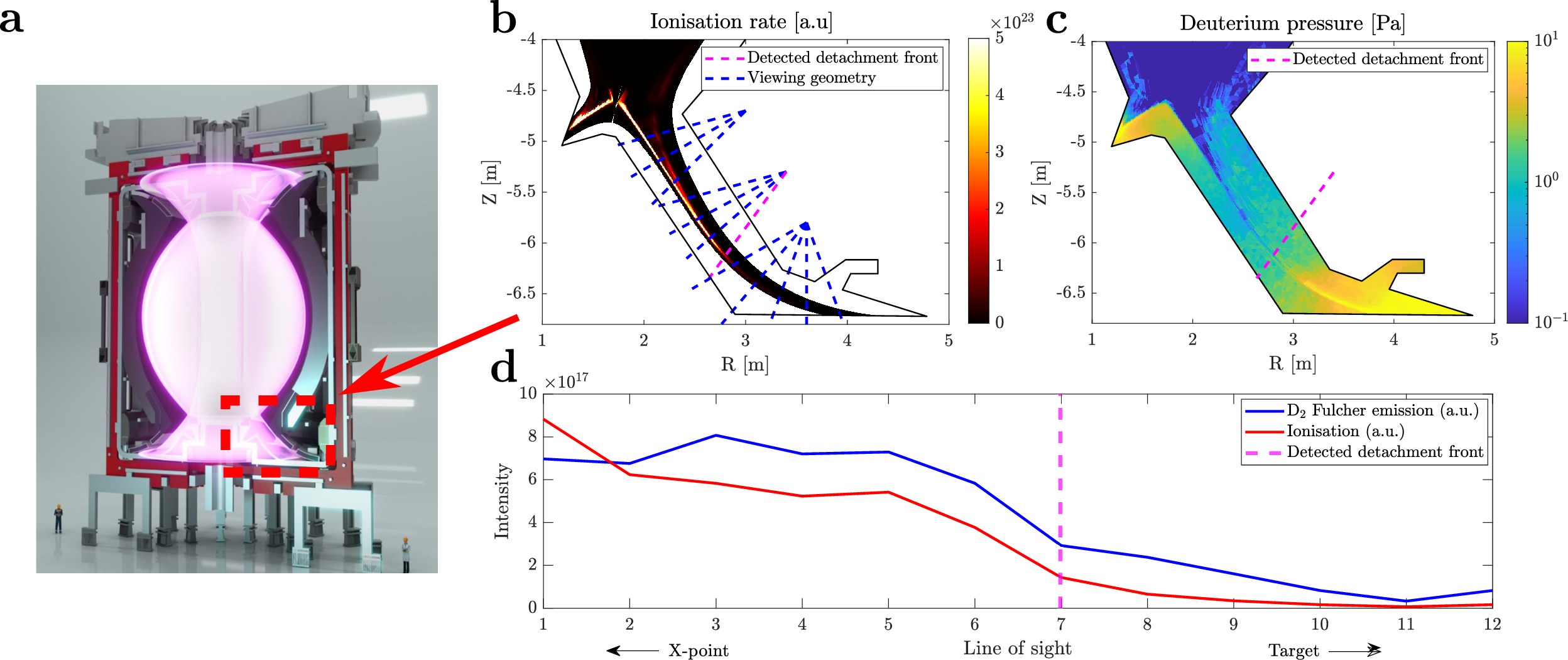}
		\caption{\textbf{Conceptual STEP divertor detachment sensor}. \textbf{a} Artistic impression of the STEP reactor\cite{Newton2022PhysicsdriversSTEPdivertorconceptdesignSTEPbaselinepowerexhaustscenario}, with the lower divertor region indicated by the red box \textbf{b} Divertor ionisation rate in STEP for a deeply detached SOLPS-ITER simulation  \cite{Osawa2023SOLPSITERanalysisproposedSTEPdoublenullgeometryimpactdegreedisconnectionpowersharing} with overlayed synthetic spectroscopy viewing geometry and detected detachment front, \textbf{c} Divertor deuterium pressure, with overlayed detected detachment front  \textbf{d} Synthetic $\mathrm{D}_2$ Fulcher band spectroscopy signals, for the sightlines shown in (b), indicating the correlation to the "true" ionisation source from the simulation and subsequent detection of the detachment front.}
		\label{fig:STEP}
	\end{figure*}


	


In this work, we have demonstrated key benefits of ADCs for exhaust control: 1) an major increase in operating window which significantly increases the ability to passively absorb transients and therefore 2) greatly improves detachment control through a resilience to re-attachment and radiative collapse. The tight baffling isolates each divertor for other operating regions which 3) enables combined control. These benefits illustrate how ADCs can be extremely beneficial for detachment control in fusion power reactors and are a viable risk mitigation strategy for power exhaust handling in fusion reactors.



A practical application where these benefits of ADCs are employed to improve reactor power exhaust is the Spherical Tokamak for Energy Production (STEP) (see methods), expected to deliver power to the UK grid in the 2040s\cite{Wilson2021STEPpathwayfusioncommercialization,Meyer2023Proceedings29thIAEAFusionEnergyConference,Muldrew2024ConceptualdesignworkflowSTEPPrototypePowerplant}. The compact design features tightly baffled long-legged divertors, with an increased strike point radius in a double-null configuration (Fig. \ref{fig:STEP}a), strongly resembling the MAST-U setup. 


Beyond illustrating the capabilities of ADCs to mitigate key risks in reactors, our research also provides practical implications for reactor detachment control. Monitoring the spatially detached regime is generally complex, requiring detailed analysis techniques that cannot be applied in real-time \cite{Verhaegh2021novelhydrogenicspectroscopictechniqueinferringroleplasmamoleculeinteractionpowerparticlebalancedetachedconditions,Perek2022spectroscopicinferenceSOLPSITERcomparisonfluxresolvededgeplasmaparametersdetachmentexperimentsTCV}. This work enables real-time detachment diagnosis, using $\mathrm{D}_2$ Fulcher band emission as a direct indicator for the detachment front location. 


We demonstrate the scientific feasibility of this sensor technique by using STEP SOLPS-ITER simulations \cite{Osawa2023SOLPSITERanalysisproposedSTEPdoublenullgeometryimpactdegreedisconnectionpowersharing} (Fig. \ref{fig:STEP}) to generate synthetic $\mathrm{D}_2$ Fulcher emission brightnesses (Fig. \ref{fig:STEP}d). As opposed to using imaging systems, we showcase that our sensor technique can even work with a synthetic spectroscopy setup with a limited, fictitious, viewing geometry (Fig. \ref{fig:STEP}b). Such an implementation can be shielded from the harsh reactor environment \cite{Luis2021NuclearanalysisDEMOdivertorsurveyvisiblehighresolutionspectrometer} and is thus more reactor relevant.  The simulation shows a highly detached lower divertor as a result of a power-sharing imbalance\cite{Osawa2023SOLPSITERanalysisproposedSTEPdoublenullgeometryimpactdegreedisconnectionpowersharing}, pushing the lower divertor ionisation front and neutral buffer upstream (Fig. \ref{fig:STEP}c). Such power-sharing imbalances are expected to be one of the most prominent disturbances in STEP \cite{Osawa2023SOLPSITERanalysisproposedSTEPdoublenullgeometryimpactdegreedisconnectionpowersharing,Lennholm2024ControllingaNewPlasmaRegime}, our results show this can be diagnosed by tracking the 50\% edge of the $\mathrm{D}_2$ Fulcher emission profile in the lower divertor. This demonstrates the scientific feasibility of the highlighted sensor technique in reactors. Consequently, it is well suited to play a central role in the collection of diagnostics which are ultimately used to diagnose the divertor for exhaust control.


The ultimate goal of an exhaust control system is to maintain feasible exhaust conditions in the presence of disturbances, requiring: 1) divertor diagnosis with suitable sensors, 2) an intrinsic capability to passively absorb transients, and 3) independent control of both divertors. Our pioneering MAST-U results show great promise in these aspects. Independent, simultaneous control of the lower/upper divertors is a major milestone planned to be demonstrated in the 2024-2025 MAST-U physics campaign. Higher external heating levels ($>10$ MW) are planned from 2026, enabling experimental validation at more reactor-relevant powers. A blend of hydrogenic and impurity gases is required for power exhaust control in reactors \cite{Lennholm2024ControllingaNewPlasmaRegime,Biel2022DevelopmentconceptbasisDEMOdiagnosticcontrolsystem} and higher power operation will facilitate studying this on MAST-U.

To conclude, our first demonstration of exhaust control in alternative divertors has highlighted major benefits for power exhaust management and its control. Therefore, implementing alternative divertors can be extremely beneficial for exhaust control and has major potential to solve the fusion exhaust problem. This research has solidified alternative divertors as a viable risk mitigation strategy towards manageable heat loads in fusion reactors.

\FloatBarrier
\begin{figure*}
	\makebox[1.1\textwidth][c]{
		\centering
		\rule{0.3\paperwidth}{0.4pt}\textit{ End of main text} \rule{0.3\paperwidth}{0.4pt}}
\end{figure*}
\FloatBarrier
\clearpage

 
\section*{Methods}\label{sec:Methods}
\subsubsection*{MAST-U fusion experiment}
The Mega Ampere Spherical Tokamak-Upgrade (MAST-U) is a tokamak fusion research experiment at the Culham Centre for Fusion Energy (CCFE) in the United Kingdom \cite{Morris2018MASTUpgradeDivertorFacilityTestBedNovelDivertorSolutions}(extended data Fig. \ref{edfig:mast-u}). It is a compact, spherical \cite{PengSPHERICALTORUSCOMPACTFUSIONLOWHELD} device with a major radius of 0.85 m and minor radius of 0.65 m, featuring a double-null design, i.e. both upper and lower divertors. MAST-U is especially constructed to explore alternative divertor configurations through its extreme divertor shaping capabilities. Notably, it is the first device designed to run in the Super-X alternative divertor configuration. Both divertor chambers are closed off from the main chamber (i.e. tightly baffled \cite{Reimerdes2021InitialTCVoperationbaffleddivertor}) to increase neutral compression, further improving exhaust performance \cite{Fishpool2013MASTupgradedivertorfacilityassessingperformancelongleggeddivertors}. 

\subsubsection*{Experimental scenario}
The experiments in this work employ the South-West Neutral Beam Injection (NBI) to increase core temperature and density, leading to increased power flowing into the divertor region. Low-confinement (L-mode) operation was used, with a plasma current of 750 kA. A deuterium main chamber gas valve located on the low-field side is used for plasma fuelling perturbations, in addition, perturbations were applied with upper and lower divertor valves for some experiments. Density control\cite{Derks2024DevelopmentrealtimedensityfeedbackcontrolMASTULmode} is employed to achieve line-averaged densities of 2-3$\cdot10^{19}$ [$\mathrm{m}^3$] before it is disabled during the perturbation or feedback-control phase of the experiments. In addition to the Conventional divertor, we employ the Elongated and Super-X alternative divertor configurations in this study.

\subsubsection*{Alternative Divertor Configurations}
Alternative divertor configurations (ADCs) are considered as an exhaust solution for future devices through their superior performance, with several, different, designs currently under investigation \cite{Verhaegh2024InvestigationsofatomicmolecularprocessesofNBI-heateddischargesintheMASTUpgradeSuper-Xdivertorwithimplicationsforreactors,Labit2017ExperimentalstudiessnowflakedivertorTCV,Militello2021PreliminaryanalysisalternativedivertorsDEMO}. These designs leverage variations in magnetic topology to enhance particle, power, and momentum losses in the divertor region, and spread the power over a larger area.  This leads to a lower plasma temperature, heat flux, and in some cases, increased access to the detached plasma regime \cite{Verhaegh2023roleplasmaatommoleculeinteractionspowerparticlebalancedetachmentMASTUpgradeSuperXdivertor}. Drawbacks of alternative designs however are, an increase in cost, engineering complexity, and a reduction in plasma volume as more space is taken up by the spatially larger divertor within the vacuum vessel \cite{Reimerdes2020AssessmentalternativedivertorconfigurationsexhaustsolutionDEMO,Militello2021PreliminaryanalysisalternativedivertorsDEMO}. The implementation of ADCs may serve as a risk mitigation strategy if conventional divertors cannot withstand the power exhaust in a reactor implementation.

One of the most prominent ADCs, and a focus in this work, is the Super-X divertor \cite{Valanju2009SuperXdivertorshighpowerdensityfusiondevices, Havlickova2015SOLPSanalysisMASTUdivertoreffectheatingpowerpumpingaccessdetachmentSuperxconfiguration,Havlickova2015effectSuperXdivertorMASTUpgradeimpurityradiationmodelledSOLPS} (Fig. \ref{fig:overview}). The strike point is placed at large major radius, leading to an increased surface area on which the power is deposited and promotes interaction with neutral particles through an increased particle path of travel (connection length) from X-point to target. The Super-X design has been demonstrated to significantly improve the target conditions, provides increased access to the detached plasma regime \cite{Verhaegh2023SpectroscopicinvestigationsdetachmentMASTUpgradeSuperXdivertor,Verhaegh2023roleplasmaatommoleculeinteractionspowerparticlebalancedetachmentMASTUpgradeSuperXdivertor}, and in this paper, significant benefits in handling fast transients.

We also consider the Elongated divertor configuration in this work (Fig. \ref{fig:overview}). This design only employs modest shaping compared to the Super-X, but has been demonstrated to already achieve significant performance gains \cite{Verhaegh2024Long-leggedDivertorsandNeutralBafflingasaSolutiontotheTokamakPowerExhaustChallenge,Verhaegh2024InvestigationsofatomicmolecularprocessesofNBI-heateddischargesintheMASTUpgradeSuper-Xdivertorwithimplicationsforreactors}.

\subsubsection*{MAST-U gas system}
The MAST-U gas system consists of a collection of piezo-electric gas valves which connect to the vacuum vessel through pipes, with various injection points in both main chamber and divertor \cite{McArdle2020MASTUpgradeplasmacontrolsystem}. Calibrations are used to convert a requested gas flow rate to a voltage which is subsequently applied to the piezo element. These calibrations are static, and hence do not take any gas valve or gas flow dynamics into account. In addition, they assume a linear voltage-to-flow rate relation which is not accurate near the closing voltage \cite{Liseli2020OverviewPiezoelectricSelfSensingActuationNanopositioningApplicationsElectricalCircuitsDisplacementForceEstimation} and does not take hysteresis of the piezo-element into account. The injected gas flow rate therefore carries considerable uncertainty \cite{Derks2024DevelopmentrealtimedensityfeedbackcontrolMASTULmode}.

\begin{figure*}
       	\centering
       \includegraphics[width=0.49\textwidth]{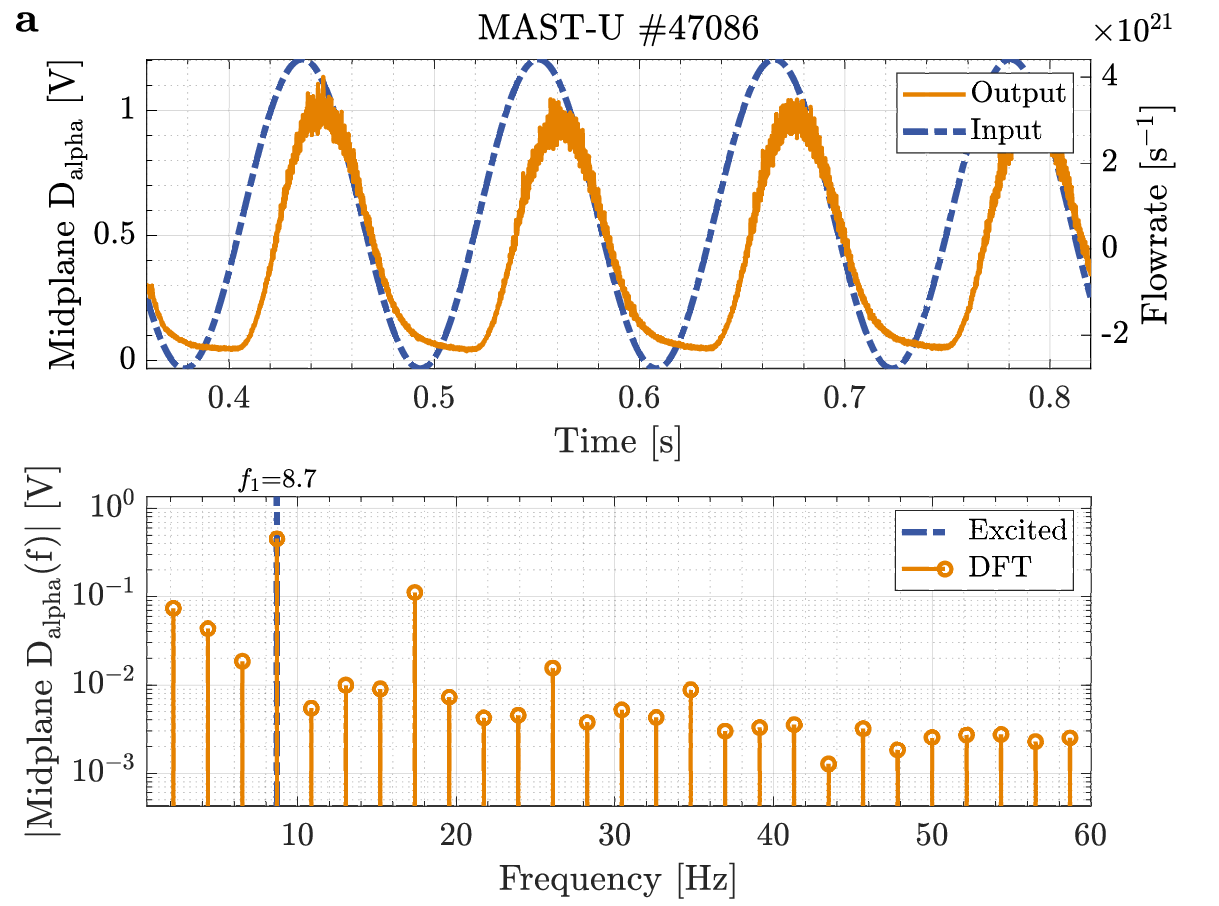}
       \includegraphics[width=0.49\textwidth]{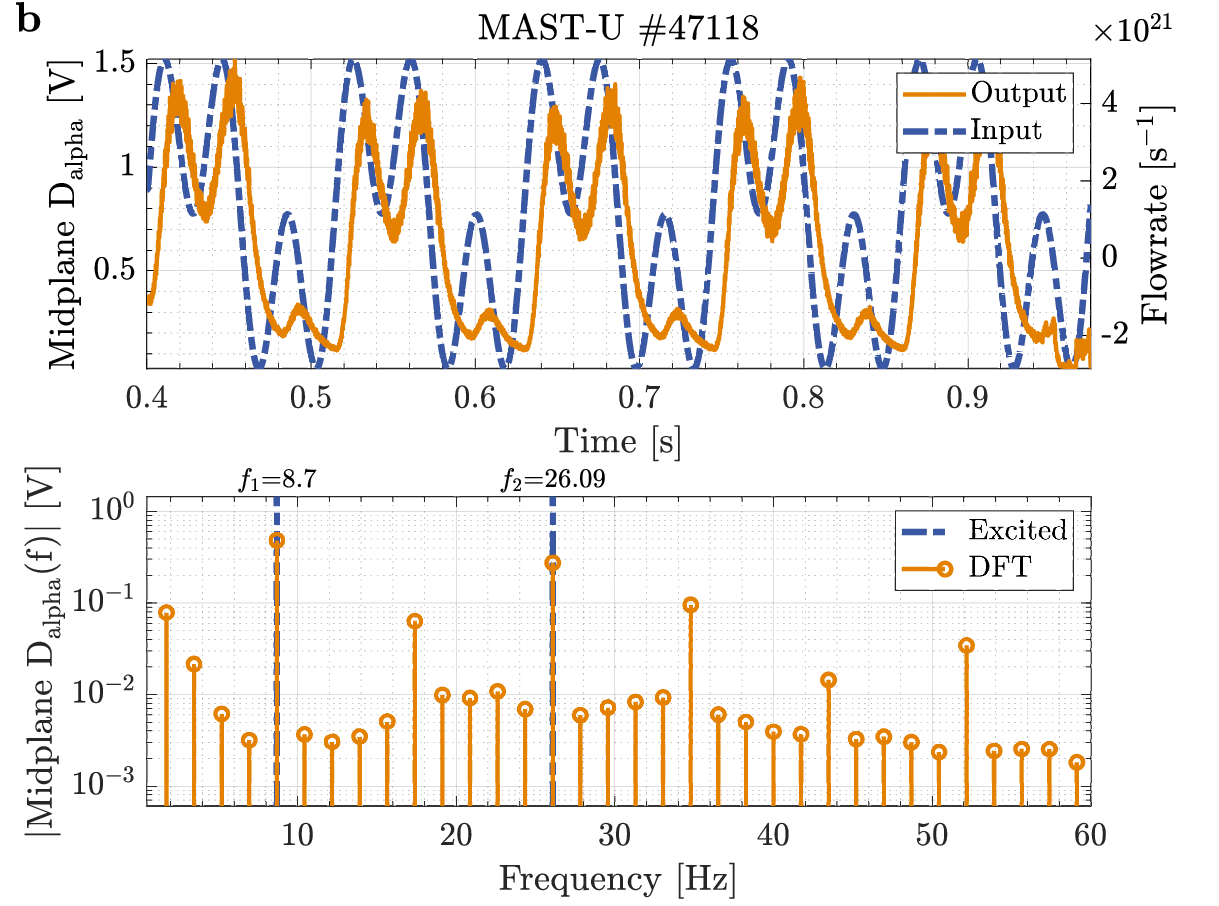}
        	\renewcommand\figurename{Extended data fig}
        	\caption{\textbf{\hspace{-0.2cm}$|$ Main chamber gas valve response} \textbf{a.} Reponse of the HM10ET midplane $\mathrm{D}_\mathrm{alpha}$ filterscope (-) to main chamber gas valve (\texttt{-{}-}) perturbations with single sine (\textbf{a}) and multisine (\textbf{b}) perturbations, both exhibiting a response on harmonics of the excited frequencies, indicative of non-linear components.}
        	\label{edfig:gasresponse}
        \end{figure*}

The main chamber valve used in these experiments (LFSD\_BOT\_L09) is positioned close to a Dalpha filterscope (HM10ET). The Dalpha intensity measured by this diagnostic represents plasma neutral interaction and can therefore serve as an indication of the gas valve response (extended data Fig. \ref{edfig:gasresponse}). During the experiments presented in this paper, the utilized flow rate was quite low, operating near the near the gas valve closing voltage, to prevent an extremely detached divertor state where the $\mathrm{D}_2$ Fulcher band is out of view of the MWI diagnostic. Therefore, we can observe some distortion at low voltages, evident as non-linear components in the frequency domain \cite{Schoukens2009}. However, these non-linear contribution do not exceed the noise floor for the front position measurement and are therefore assumed to not significantly influence the analysis (Fig. \ref{fig:lines}).

Dalpha filterscopes positioned in the upper (HU10SXDT) and lower divertors (HL02SXDT) are employed to also check the functioning of the respective divertor valves (LFSD\_TOP\_U0102, LFSD\_BOT\_L0506). These signals are carry more noise compared to the main chamber signals since the divertor filterscopes are positioned further away from the valves, nevertheless, they clearly indicate the divertor valve perturbations (Fig \ref{fig:interaction lower}, \ref{fig:interaction upper}).

\subsubsection*{ $\mathrm{D}_2$  Fulcher band emission}
Throughout this work, we use  $\mathrm{D}_2$ Fulcher band emission to diagnose the divertor conditions.  $\mathrm{D}_2$  Fulcher band emission originates from electronically excited molecules which requires similar plasma temperatures to ionisation. Therefore, Fulcher band emission has been presented as a quantitative method to infer the ionisation rate and position \cite{Osborne2023InitialFulcherbandobservationshighresolutionspectroscopyMASTUdivertor,Verhaegh2023SpectroscopicinvestigationsdetachmentMASTUpgradeSuperXdivertor,Verhaegh2021novelhydrogenicspectroscopictechniqueinferringroleplasmamoleculeinteractionpowerparticlebalancedetachedconditions}. The steep temperature dependence creates a clearly defined front position that can be tracked relatively easily using filtered imaging or spectroscopy. The position of this ionisation region is a fundamental indicator of the divertor detachment state \cite{Verhaegh2023roleplasmaatommoleculeinteractionspowerparticlebalancedetachmentMASTUpgradeSuperXdivertor,Perek2022spectroscopicinferenceSOLPSITERcomparisonfluxresolvededgeplasmaparametersdetachmentexperimentsTCV}. Contrary to impurity emission fronts which are routinely used to diagnose detachment \cite{Fevrier2021a,RavensbergenControl2021,Koenders2023ModelbasedimpurityemissionfrontcontrolusingdeuteriumfuelingnitrogenseedingTCV}, Fulcher band emission is unaffected by impurity transport. This results in a more reliable, machine-independent, and direct indication of the divertor detachment state. 

\subsubsection*{Inversion-based front tracking}
Tracking of the Fulcher band emission front is achieved using spectrally filtered images from the Multi-Wavelength-Imaging camera system \cite{Feng2021Development11channelmultiwavelengthimagingdiagnosticdivertorplasmasMASTUpgrade,Perek2019MANTISrealtimequantitativemultispectralimagingsystemfusionplasmas}, positioned in the lower divertor of MAST-U (Fig. \ref{fig:overview}b, \ref{fig:feedback}a,f). The raw camera images from the Fulcher band-filtered channel (Fig \ref{fig:feedback}a,f) are inverted to achieve a poloidal emissivity profile (Fig. \ref{fig:interaction lower}c, \ref{fig:interaction upper}c). The front position is then taken as the position with 50\% extinction from the maximum intensity \cite{Wijkamp2023CharacterisationdetachmentMASTUSuperXdivertorusingmultiwavelengthimaging2Datomicmolecularemissionprocesses}. The mission front tracking algorithm outputs $\mathrm{L}_{\mathrm{tar}}$ and $\mathrm{L}_{\mathrm{X}}$, the distance from the emission front to the divertor target and X-point, measured along the divertor leg in the poloidal plane (Fig. \ref{fig:SXDvsECD}d).

The Conventional divertor leg is largely out of view of the MWI diagnostic. Therefore, we employ the newly available X-point Imaging System (XPI) to achieve front tracking for the Conventional divertor in a similar manner. This system was not yet available in the earlier Super-X and Elongated divertor midplane fuelled system identification and feedback control shots. Combined XPI and MWI inversions are performed for the tracking of the divertor fuelled system identification experiments (Fig. \ref{fig:interaction lower}, \ref{fig:interaction upper}c).

The MWI, XPI or combined inversions required for this front-tracking method are computationally expensive, hence, front tracking using inverted images is currently only available offline, i.e. after the experiment has completed \cite{Wijkamp2023CharacterisationdetachmentMASTUSuperXdivertorusingmultiwavelengthimaging2Datomicmolecularemissionprocesses}. Machine learning-based acceleration techniques have shown promising results\cite{vanLeeuwen2022MScthesisMachineLearningAcceleratedTomographicReconstructionMultispectralImagingTCV} and might allow for a future extension to real-time operation. In this paper, we employ inversion-based front tracking only for offline analysis of system identification experiments and we rely on a different routine for real-time front tracking.

\subsubsection*{Real-time front tracking}
Real-time emission front tracking is achieved using raw, un-inverted camera images directly through a dedicated algorithm first employed in the TCV tokamak \cite{Ravensbergen2020Developmentrealtimealgorithmdetectiondivertordetachmentradiationfrontusingmultispectralimaging}. A dedicated fast coordinate transformation\cite{Hommen2014Opticalboundaryreconstructionshapecontroltokamakplasmas} is used to achieve the transform from raw camera images to the poloidal plan without requiring camera inversion. Although real-time capable, the coordinate transformation introduces additional noise. We take the front as the 50\% extinction from the maximum intensity along the leg, identical to the inversion-based front tracking routine. In absence of a real-time magnetic equilibrium reconstruction, we prescribe the divertor leg position a priori such that the distance to target along the divertor leg $\mathrm{L}_{\mathrm{tar}}$ can be calculated (Fig. \ref{fig:feedback}a,f). The full front tracking algorithm is executed within the 2.5 ms time window of the MWI diagnostic, allowing for 400 Hz operation. The computed front position is fed through an optical-analogue connection into the plasma control system\cite{McArdle2020MASTUpgradeplasmacontrolsystem} where the exhaust controller is located.

The use of the fast coordinate transformation \cite{Hommen2014Opticalboundaryreconstructionshapecontroltokamakplasmas} introduces a requirement for tangential sightlines. A camera system records a two dimensional projection of a three dimensional feature, implying that each individual pixel in the camera image consists of the line integrated emission along its sightline. Pixels corresponding to a sightline that is tangent to the light-emitting divertor leg will show a peak in intensity compared to neighbouring sightlines that either intersect or miss the leg\cite{Hommen2014Opticalboundaryreconstructionshapecontroltokamakplasmas}. Therefore, only the sightlines that are tangential to the divertor emission result in a recognisable divertor leg on the raw camera image.

\begin{figure*}
    \centering
    \includegraphics[width=\linewidth,trim=0.8cm 0 0 0, clip]{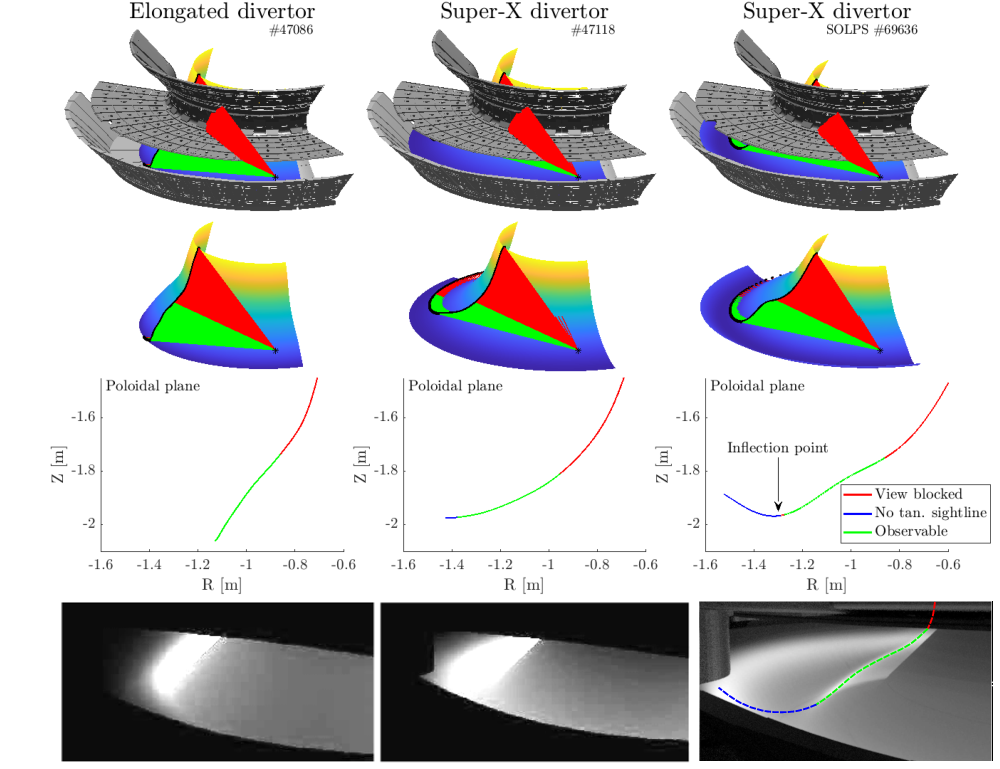}
    	\renewcommand\figurename{Extended data fig}
    \caption{\textbf{\hspace{-0.2cm}$|$ Geometric analysis for the MAST-U Super-X divertor}, indicating how the coordinate transformation used in the real-time tracking algorithm will fail close to the inflection point. This figure will be remade}
    \label{edfig:geoanalysis}
\end{figure*}

The observable region for real-time front tracking can be quantified through a dedicated geometric analysis. We consider both the Elongated and Super-X divertor geometries used in the experiments presented in this work as well as a Super-X divertor variant previously presented in Simulations$^{[?]}$ (extended data Fig. \ref{edfig:geoanalysis}). The core assumption of this analysis is that the divertor emission is located on the magnetic divertor leg. The camera position is inferred from spatial calibrations obtained through the Calcam\cite{Silburn2020CALCAMsoftwarepackageversion} software package. In Elongated geometry, the full leg can be observed from divertor baffle to target. However, for the Super-X divertor geometries, no tangential sightline exists near the target. In the poloidal plane, a negative $\frac{\mathrm{dR}}{\mathrm{dZ}}$ is required for a tangential point to exist, otherwise, a sightline will intersect with the plasma by definition. We designate the point where $\frac{\mathrm{dR}}{\mathrm{dZ}}$ changes sign as the inflection point. Its location greatly influences the observable region near the target, evident by comparing the two Super-X geometry variants (extended data Fig. \ref{edfig:geoanalysis} e-l). We therefore conclude that the observable region for emission front tracking in MAST-U using raw camera images from the MWI diagnostic is bound by the baffle at the divertor entrance and the inflection point near the divertor target.

In the experiments presented in this work, the Super-X divertor is always in a detached state, i.e. the Fulcher band emission is far removed from the target. Coupled with an inflection point location close to the target, we conclude that the lower limit for real-time front tracking has likely not influenced these experiments. Nevertheless, the real-time front tracking range can be severely limited for other Super-X divertor geometries, especially in more attached conditions.

\subsubsection*{System Identification}
We experimentally identify the exhaust dynamics in MAST-U through system identification. This method relies on observing the system response to applied perturbations and has been employed successfully to identify the exhaust dynamics on several devices \cite{RavensbergenControl2021,Koenders2022SystematicextractioncontrolorientedmodelperturbativeexperimentsSOLPSITERemissionfrontcontrolTCV,Koenders2023ModelbasedimpurityemissionfrontcontrolusingdeuteriumfuelingnitrogenseedingTCV}. In addition to allowing for the \textit{design} of a feedback controller, the experimental identification of exhaust dynamics supports the development of physics-based dynamic models to inform control system design for future devices.

The considered dynamic system (Fig. \ref{fig:overview}b) includes the gas system, plasma response, and MWI sensor dynamics. Note that the dynamics of the piezo-electric gas valve and its associated piping is included in the system (see MAST-U gas system section).  The system input is the requested flow rate $\mathrm{u}$, the output is the Fulcher band poloidal emission front position $\mathrm{L}_{\mathrm{tar}}$. 

We perturb the system input with especially designed signals, consisting of a single sine or a sum of sinusoidal signals. This allows the signal power to be focussed on specific frequencies of interest. We use only a few frequencies per experiment, driven by the low Signal-to-Noise Ratios (SNR) generally observed in detachment measurements\cite{RavensbergenControl2021}. The available time for perturbation is only 300-400 ms due to the relatively short \textless1 s total shot duration in MAST-U. We require at least three periods per frequency to generate errorbars on the data, the lowest frequency $f_1$ within the measurement window is therefore around 8-10 Hz. The upper limit is set by the gas system, above this limit the gas system will no longer follow the prescribed perturbation signal, taken as roughly 50 Hz \cite{Koenders2022SystematicextractioncontrolorientedmodelperturbativeexperimentsSOLPSITERemissionfrontcontrolTCV}.

The perturbation signals consist of integer multiples, or harmonics, ($f_3$, $f_5$, $f_{...}$ ) of the ground frequency ($f_1$). Generally, only odd frequency components are excited to observe possible quadratic non-linear effects\cite{Schoukens2009,VanBerkel2020CorrectingnonperiodicbehaviourperturbativeexperimentsApplicationheatpulsepropagationmodulatedgaspuffexperiments}. However, this quickly drives us towards frequencies above the stipulated 50 Hz gas system limit. Therefore, we occasionally opt to use $f_1$ and $f_2$ or perturb only a single frequency at the expense of experimental time.

\begin{figure*}
	
		\centering
		\includegraphics[width=0.49\textwidth]{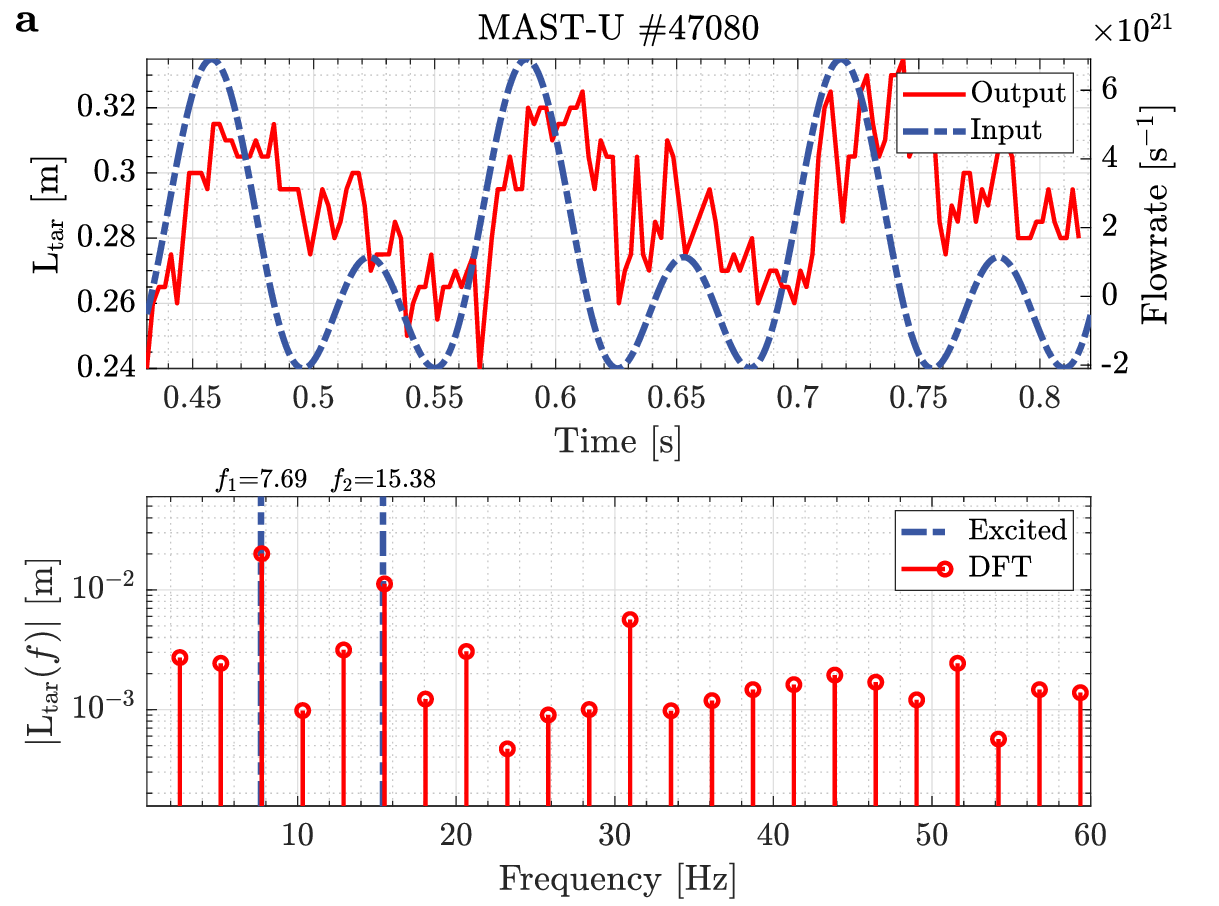}
		\includegraphics[width=0.49\textwidth]{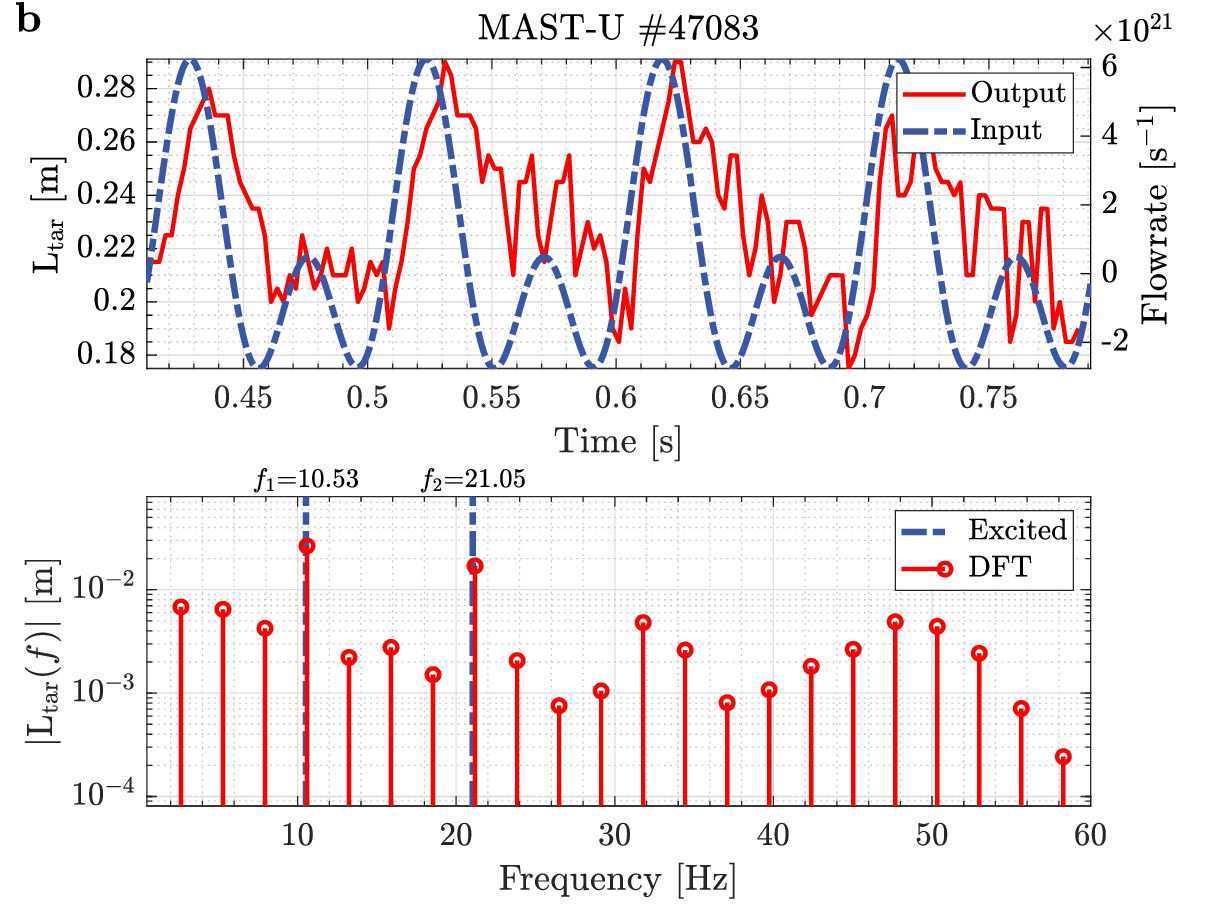}
		\includegraphics[width=0.49\textwidth]{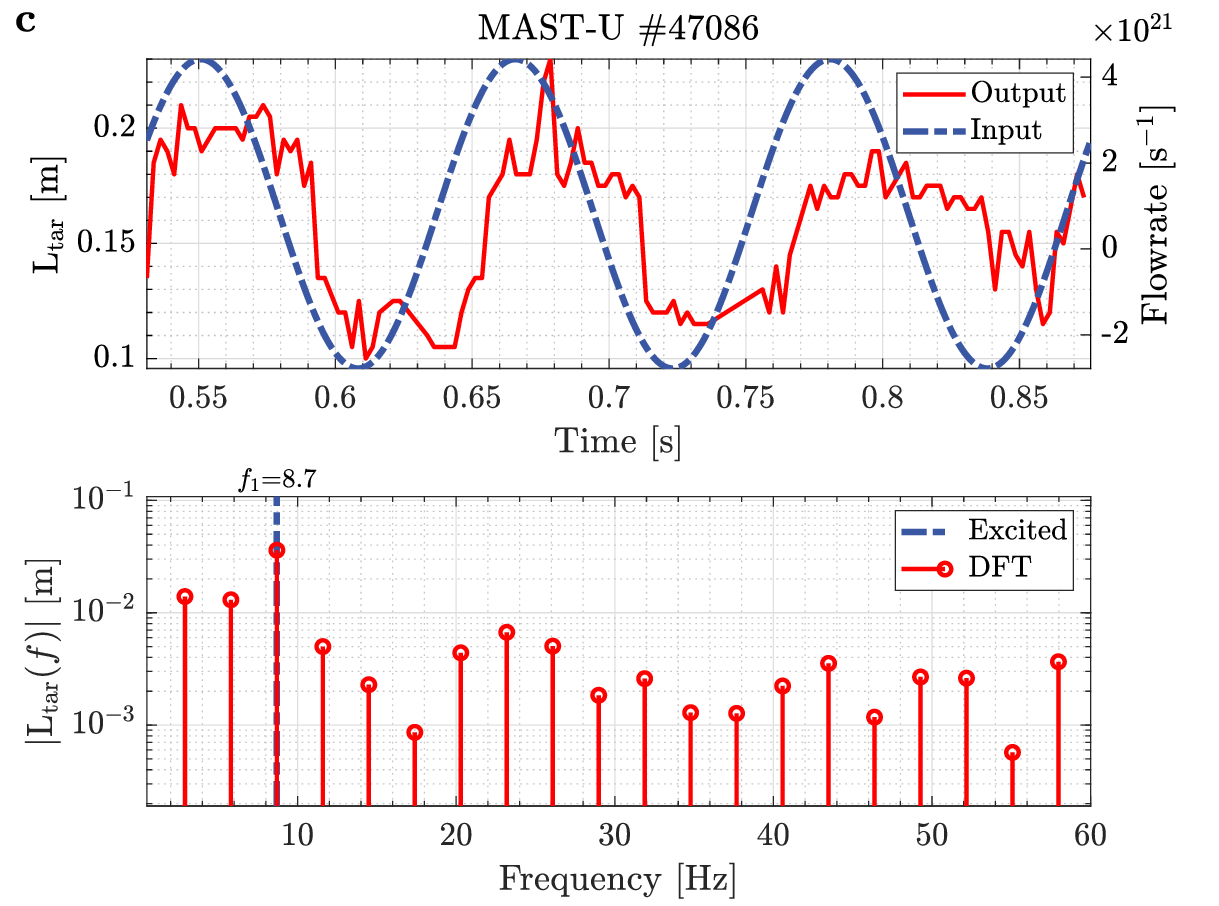}
		\includegraphics[width=0.49\textwidth]{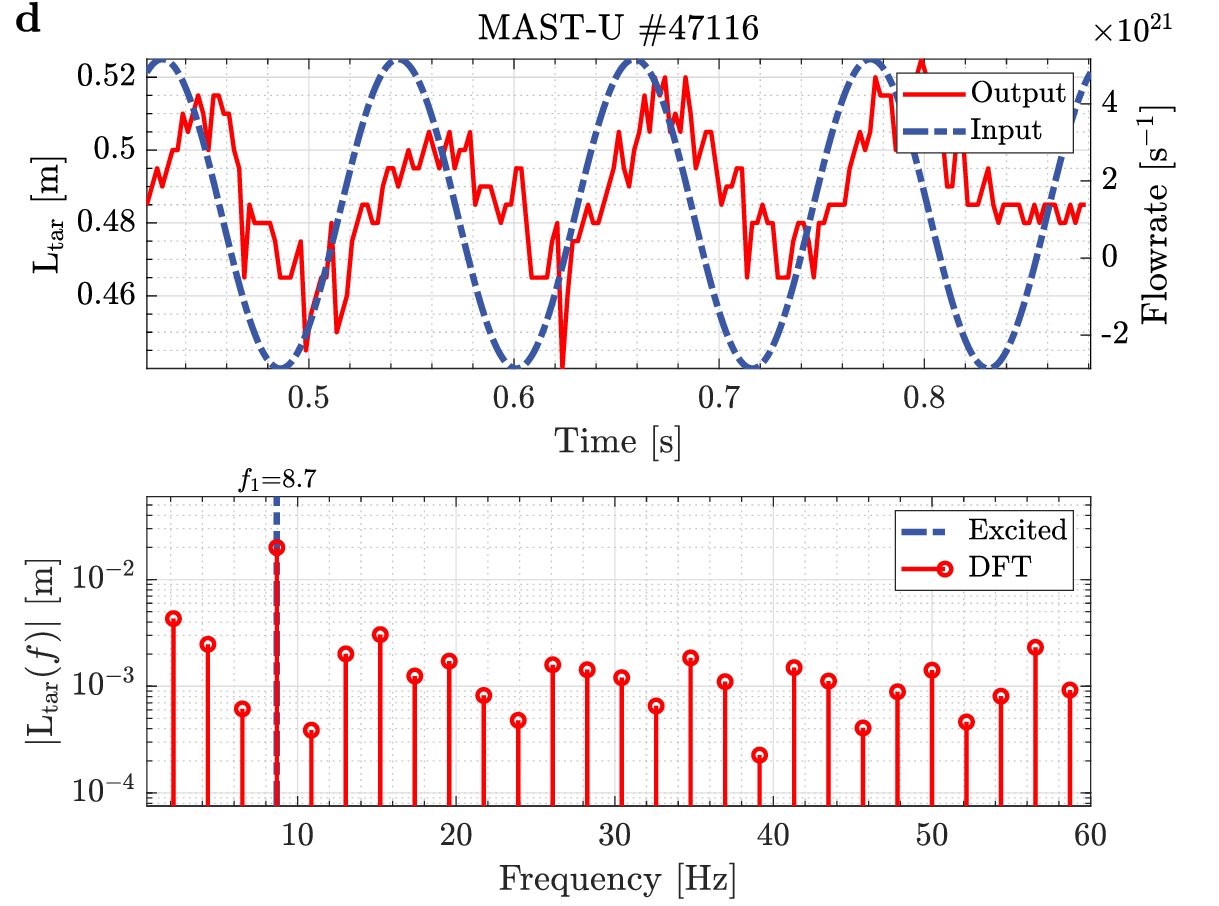}
		\includegraphics[width=0.49\textwidth]{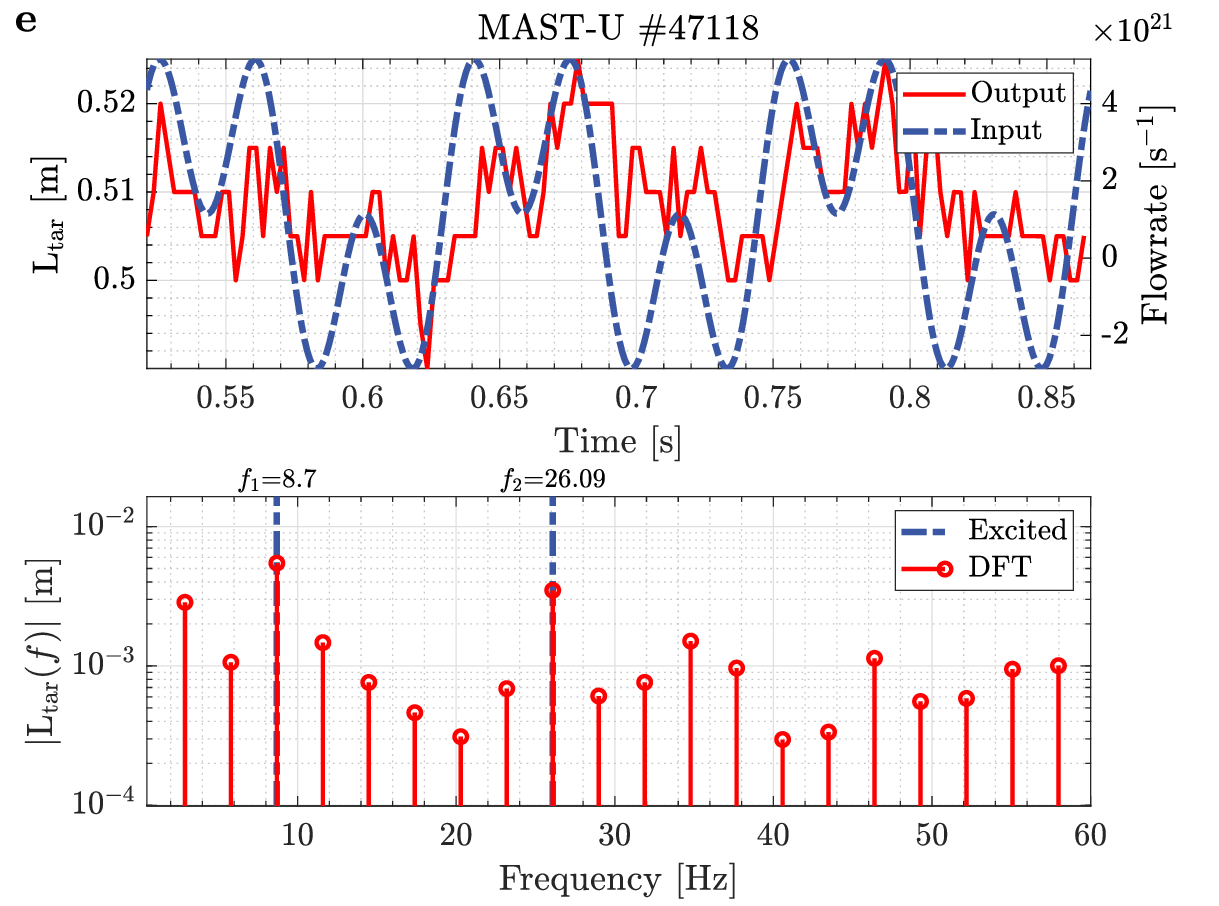}
		\includegraphics[width=0.49\textwidth]{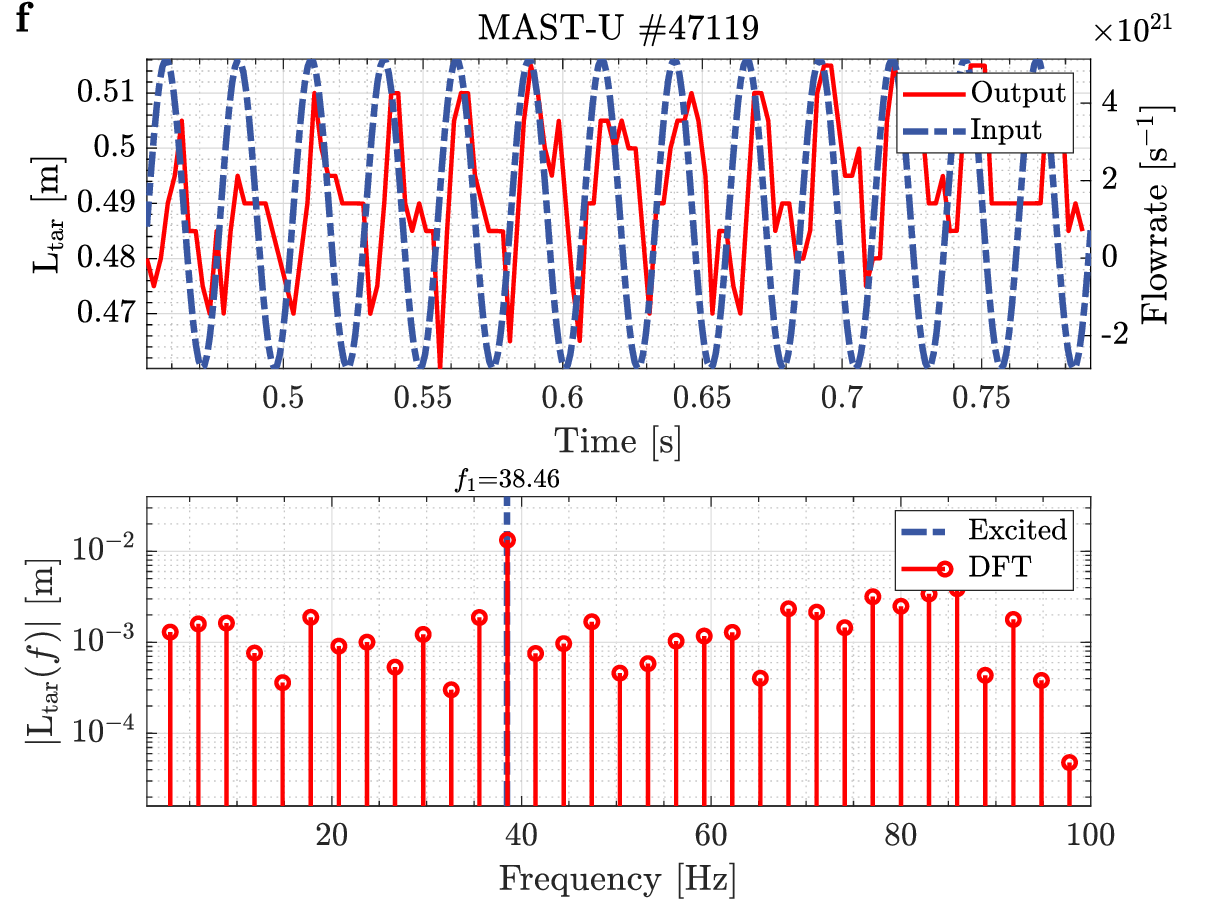}
	\renewcommand\figurename{Extended data fig}
\caption{\textbf{\hspace{-0.2cm} $|$ System identification in MAST-U} Observed $\mathrm{D}_2$ Fulcher band front position $\mathrm{L}_{\mathrm{pol}}$  (-) response to gas valve flow request perturbations (\texttt{-{}-}) in Elongated (\#47080, \#47083, \#47086) and Super-X (\#47116, \#47118, \#47119) divertor geometry. In time (top) and frequency domain (bottom). The output response is clearly above the noise level for the exited frequencies while remaining below the noise level for the odd integer multiples, indicating dominantly linear dynamics\cite{Schoukens2009}.}
	\label{edfig:frontresponse}
\end{figure*}

The input and output signals are transformed from the time domain to the frequency domain using the Discrete-Fourier Transform (DFT). We identify the Frequency Response Function (FRF) of the system by dividing the observed output over input in frequency domain. The Local Polynomial Method (LPM) is used\cite{Schoukens2009,VanBerkel2020CorrectingnonperiodicbehaviourperturbativeexperimentsApplicationheatpulsepropagationmodulatedgaspuffexperiments} to correct for transient effects. The FRF is a local linearisation of the input-output dynamics, non-linear effects will not be captured by this method. This is standard practice in control theory and applicable for this system since the DFT response contains predominantly the excited frequencies (Fig. \ref{fig:lines}, extended data fig. \ref{edfig:frontresponse}), indicating dominantly linear dynamics\cite{Schoukens2009} for the considered frequency range.

\subsubsection*{STEP fusion power reactor}
\begin{figure*}
	\hspace{-2cm}
	\makebox[1.1\textwidth][c]{
		\centering
		\includegraphics[width=0.9\textwidth]{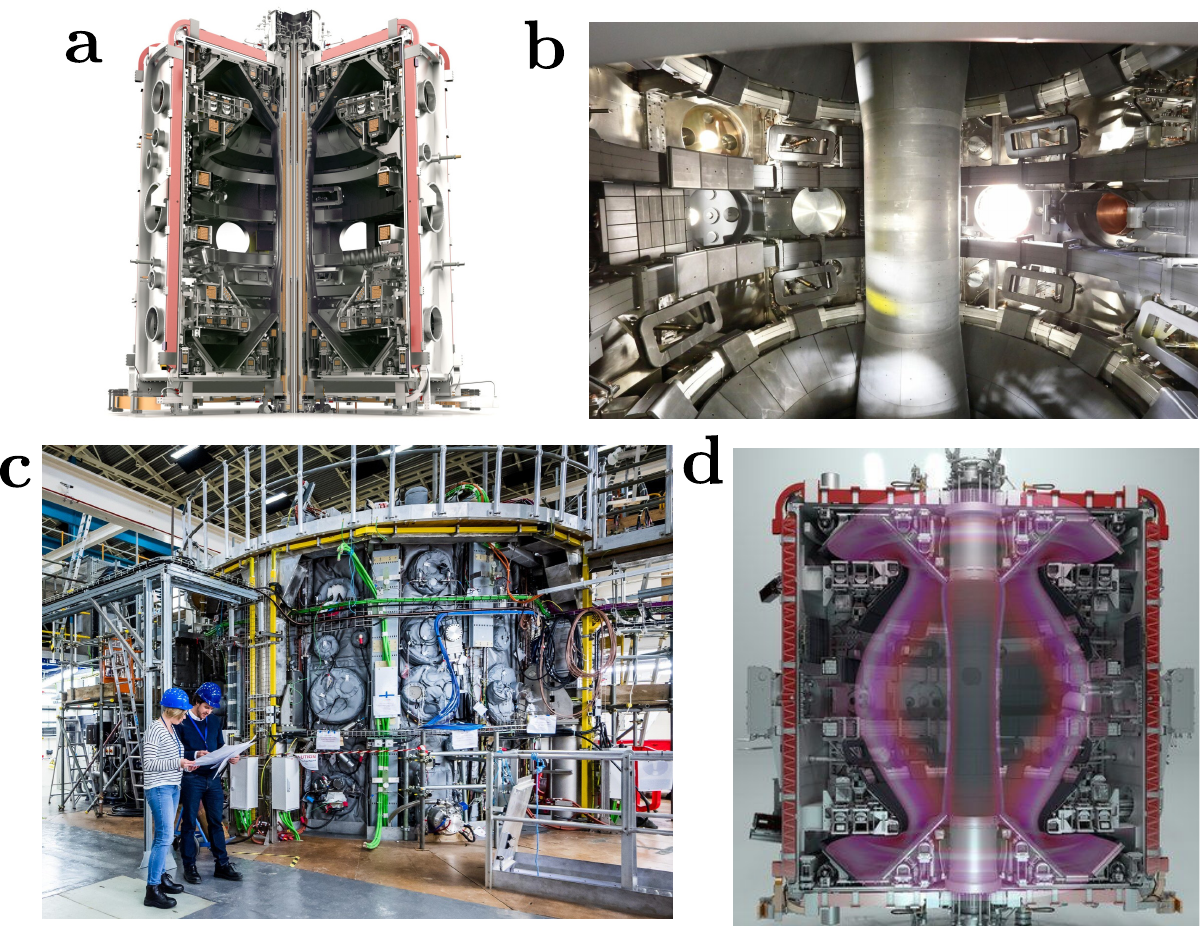}}
		\renewcommand\figurename{Extended data fig}
	\caption{\textbf{\hspace{-0.2cm}$|$ The MAST-U tokamak} \textbf{a.} Cross section of a CAD render of the MAST-U without plasma \cite{MAST-Uimagegallery}. \textbf{b,c} Pictures of the interior and exterior of MAST-U\cite{MAST-Uimagegallery} \textbf{d} Cross section of a CAD render of MAST-U with plasma, highlighting the Super-X divertor chambers \cite{MAST-Uimagegallery}. }
	\label{edfig:mast-u}
\end{figure*}

\begin{figure*}
	\hspace{-2cm}
	\makebox[1.1\textwidth][c]{
		\centering
		\includegraphics[width=0.6\textwidth]{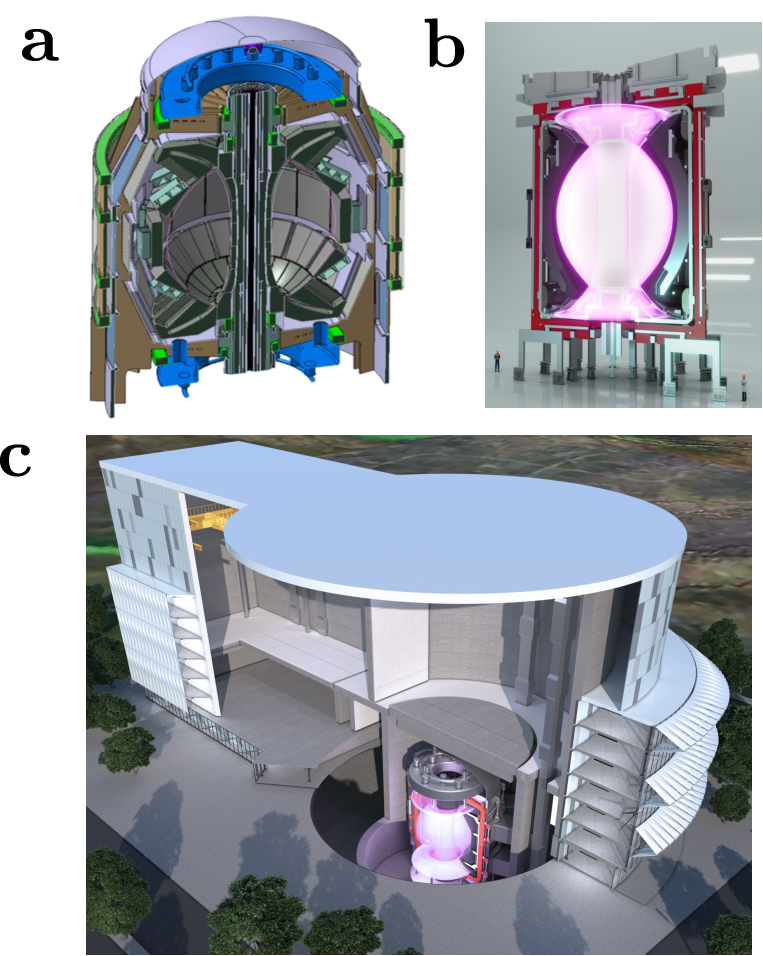}}
	\renewcommand\figurename{Extended data fig}
	\caption{\textbf{\hspace{-0.2cm}$|$ The STEP tokamak} \textbf{a.} Cross section of a CAD model of the STEP concept design without plasma \cite{McClementsSphericalTokamakEnergyProductionSTEPSteadyStateFusionReactor}. \textbf{b,c} Artistic impressions of the STEP reactor \cite{Newton2022PhysicsdriversSTEPdivertorconceptdesignSTEPbaselinepowerexhaustscenario} and its surrounding buildings \cite{McClementsSphericalTokamakEnergyProductionSTEPSteadyStateFusionReactor}. }
	\label{edfig:STEP_overview}
\end{figure*}

The Spherical Tokamak For Energy Production (STEP) is tokamak currently in the concept design phase (extended data Fig. \ref{edfig:STEP_overview}). It is a highly ambitious programme, targeting completion around 2040 with the ultimate aim of delivering fusion power to the UK grid\cite{Wilson2021STEPpathwayfusioncommercialization,Meyer2023Proceedings29thIAEAFusionEnergyConference,Muldrew2024ConceptualdesignworkflowSTEPPrototypePowerplant}. The current design has a major radius of 3.6 m and is currently expected to deliver 120 MW electrical power\cite{Wilson2021STEPpathwayfusioncommercialization,Meyer2023Proceedings29thIAEAFusionEnergyConference,Muldrew2024ConceptualdesignworkflowSTEPPrototypePowerplant}. STEP is a spherical tokamak, equipped with two tightly baffled \cite{Reimerdes2021InitialTCVoperationbaffleddivertor} divertor chambers which facilitate long-legged divertor configurations, akin to MAST-U (extended data Fig. \ref{edfig:mast-u}). Although it shares a similar design philosophy to MAST-U, its design is fundamentally differently in coping with the neutral irradiation, power cycle, tritium fuel cycle, and other requirements placed on a fusion power reactor \cite{Muldrew2024ConceptualdesignworkflowSTEPPrototypePowerplant}. The requirement of an exhaust control system for STEP to ensure manageable heat loads is a core driver for the work presented in this paper.



\section*{Acknowledgments}




This work has been carried out within the framework of the EUROfusion Consortium, funded by the European Union via the Euratom Research and Training Programme (Grant Agreement No 101052200 - EUROfusion) and from the EPSRC [grant numbers EP/W006839/1, EP/S022430/1, EP/T012250/1 and EP/N023846/1]. The Swiss contribution to this work has been funded by the Swiss State Secretariat for Education, Research and Innovation (SERI). Views and opinions expressed are however those of the author(s) only and do not necessarily reflect those of the European Union, or the European Commission or SERI. Neither the European Union nor the European Commission nor SERI can be held responsible for them.

\section*{Extended data}
\FloatBarrier
%




%


\pagebreak
\FloatBarrier
\bibliography{bibliography}

\end{document}